\documentclass[preprint,showpacs,titlepage,aps,prd,tightenlines,amsmath,byrevtex,nofootinbib]{revtex4}

\usepackage{graphicx}
\usepackage[dvips]{color}
\usepackage{epsfig}

\newcommand{\be}{\begin{equation}}
\newcommand{\ee}{\end{equation}}
\newcommand{\bea}{\begin{eqnarray}}
\newcommand{\eea}{\end{eqnarray}}
\newcommand{\beq}{\begin{equation}}
\newcommand{\eeq}{\end{equation}}

\def\anue{{\bar\nu_e}}

\def\lsim{\ ^<\llap{$_\sim$}\ }

\def\beq{\begin{equation}}
\def\eeq{\end{equation}}
\def\beqn{\begin{eqnarray}}
\def\eeqn{\end{eqnarray}}

\def\sinW2{\sin^2\theta_W}
\begin{document}

{\hbox{hep-ph/0611194}}

\baselineskip 18pt
\def\today{\ifcase\month\or
 January\or February\or March\or April\or May\or June\or
 July\or August\or September\or October\or November\or December\fi
 \space\number\day, \number\year}
\def\thebibliography#1{\section*{References\markboth
 {References}{References}}\list
 {[\arabic{enumi}]}{\settowidth\labelwidth{[#1]}
 \leftmargin\labelwidth
 \advance\leftmargin\labelsep
 \usecounter{enumi}}
 \def\newblock{\hskip .11em plus .33em minus .07em}
 \sloppy
 \sfcode`\.=1000\relax}
\let\endthebibliography=\endlist
%


\title{Determining Neutrino and Supernova Parameters with a 
Galactic Supernova}

\vskip 1.5 true cm

\renewcommand{\thefootnote}

{\fnsymbol{footnote}}

\author{Solveig Skadhauge\email{E-mail: solveig@fma.if.usp.br}}
\author{Renata Zukanovich Funchal\email{E-mail: zukanov@if.usp.br}}
\affiliation{Instituto de F\'{\i}sica, Universidade de S\~{a}o Paulo, 
 C.\ P.\ 66.318, 05315-970 S\~{a}o Paulo, SP, Brazil}


\vskip 1.5 true cm
\begin{abstract}
  We investigate the sensitivity of some of the proposed
  next-generation neutrino experiments to a galactic supernova. 
  In particular, we study how well the supernova parameters 
  (the average energies and luminosities) can be separated 
  from the unknown neutrino oscillation parameters ($\theta_{13}$ 
  and the neutrino mass hierarchy).  Three types of experiments, 
  all in the 100 kilo-ton class, are compared.  
  These are: a 540 kton water-Cherenkov detector, a 100
  kton liquid Argon detector and a 50 kton scintillator detector.  We
  demonstrate that practically all of these proposed detectors have
  the possibility to determine the hierarchy of the neutrino masses if
  the angle $\theta_{13}$ is sufficiently large
  ($\sin^2(\theta_{13}) \stackrel{>}{\sim} 10^{-4}$) and the hierarchy
  of the average energies is larger than about 20\%. They can at the same
  time determine some of the supernova parameters well. The average
  energy of the $\nu_\mu$ and $\nu_\tau$ species can be determined within  
  5\% uncertainty in most of the parameter space suggested by supernova 
  simulations.
  The detection of several separable channels measuring different
  combinations of charged current and neutral current processes is
  crucial for determining the value of $\theta_{13}$ and the
  hierarchy. However, there are cases where a few of the SN parameters 
  can be determined rather well even if only the main charged current 
  detection channel is available.
\end{abstract}

\pacs{14.60.Pq,25.30.Pt,97.60.Bw}

\maketitle
\section{\label{sec:intro} Introduction}

The fortuitous observation of a handful of neutrinos from the
supernova 1987A \cite{Hirata:1987hu,Bionta:1987qt} started the era of
experimental supernova (SN) neutrino physics.  It confirmed our main
ideas of the physics of a supernova from a core collapse, although
also a set of (minor) disagreements were found. However, the low
statistics of neutrino events collected from SN1987A, partially due to
the fact that it happened at a distance of roughly 50 kpc from Earth,
was not sufficient to really extract much information.  Evidently,
with a larger detector and a closer by supernova, the prospects for
gaining high accuracy information about both supernova and neutrino
physics are immense. In the case of a galactic SN, one can observe of
the order of 500 events per kton of detector material and there are
several proposals for neutrino detectors in the 100 kton
range~\cite{ScholbergTalk}.

A type II supernova is the death of a giant star and the huge emission
of light as well as neutrinos is an effect of the gravitational core 
collapse of the star. In fact, about 99\% of the total energy is
emitted in the form of neutrinos and antineutrinos in a roughly 10 second
interval. The consecutive burning of different elements, structured
in an onion shell form, ends at the silicon burning phase.  This phase
produces an iron core and as fusion of iron is impossible, the
gravitational collapse of the core is triggered once it reaches a
well-known size, related to the Chandrasekhar limit.  The core heats
up during the implosion and photo-disintegration of the iron atoms
starts, having as a byproduct free protons and neutrons. The
subsequent neutronization gives rise to a $\nu_e$ flux from the
deleptonization process $e+p \rightarrow \nu_e + n$.  The SN
densities in the interior of the core are so high that even neutrinos
are trapped. However, neutrinos, interacting only weakly, manage to
escape near the surface of the core region.  A solid feature in all SN
simulations is the formation of a shock wave caused by the rebounce of
material falling on to the core.  It is believed that this shock wave
will cause the explosion of the outer material of the star.  However,
it has for a long time been a pending embarrassment, that SN
simulations are incapable of producing the SN explosion as the shock
wave looses too much energy to the media and halts before erupting the
outer parts of the star.  It has been suggested that neutrino physics
might play a crucial role in re-energizing the shock
wave~\cite{bethe-wilson}, but until now a successful calculation has
not been performed.  After this so-called collapse phase, where a
neutron star or, in rare cases, a black hole is formed, there is an
accretion phase and a cooling phase.

In this article we will study the neutrino flux from the cooling
phase. It is probably the phase where the physics is best controlled,
although many things are still unknown.  The energy spectra will be
almost thermal, as the densities in the SN are so high that in fact
the neutrinos will be in thermal equilibrium.  But what are the
average energies of each neutrino flavor?  Again, a robust feature
seems to be the hierarchy of these average energies. Electron
neutrinos can interact in the medium both through charged current (CC)
and neutral current (NC) exchanges, whereas $\nu_\mu$ and $\nu_\tau$
suffer only NC interactions as there are no muons or taus in the SN
material.  As a consequence, electron neutrinos will maintain their
thermal equilibrium to a larger radius, therefore escaping with a
smaller temperature or, equivalently, lower average energy $\langle
E_e\rangle$. Furthermore, as $\nu_\mu$ and $\nu_\tau$, both, only
interact via NC, their energy spectra will have identical properties.
We will treat them as indistinguishable, denoting them with the common
index $x$.  Anti-electron neutrinos also interact via CC reactions but
with a smaller cross-section (since there are less protons than
neutrons) resulting in a higher temperature (average energy 
$\langle E_{\bar e}\rangle$) than electron neutrinos. 
So, a hierarchy of the form $\langle E_{e} \rangle < \langle E_{\bar e} 
\rangle < \langle E_{x} \rangle$ is predicted to exist. However,
the exact values of the average energies as well as the strength of
the hierarchies, vary quite substantially in different simulations and
also depend on the type of the progenitor star~\cite{garching,livermore}.

Neutrino flavor transitions have been observed in atmospheric,
solar, reactor and accelerator neutrino experiments. 
The simplest and most widely accepted way to explain these 
transitions is to allow neutrinos to have masses and mixings.
Although, the neutrino oscillation parameters have been
determined to increasingly astonishing precision during the past few
years~\cite{atm, solar,kamland,k2k,MINOS}, currently,
\begin{eqnarray}
+7.3 \times 10^{-5} {\rm eV^2} < 
& \Delta m^2_{21} & 
< +9.0 \times 10^{-5} {\rm eV^2} \nonumber \\
0.25 < & \sin^2 \theta_{12}& < 0.37 
\label{solarpar}  \\
1.5 \times 10^{-3} {\rm eV^2} < 
& \vert \Delta m^2_{32} \vert & 
< 3.4 \times 10^{-3} {\rm eV^2} \nonumber \\
0.36 < & \sin^2 \theta_{23} & \leq 0.64  
\label{eq:param}
\end{eqnarray}
at 90\% CL, some important points remain unknown.  We still lack
information on the absolute neutrino mass scale $m_0$, only an upper
bound $m_0 \lsim 0.2-0.7$ eV~\cite{wmap} exist. The neutrino mass
pattern is not yet completely established: we do not know if nature
prefers the normal ($m_3 > m_2 > m_1$) or the inverted ($m_2 > m_1 >
m_3$) mass hierarchy, where $m_1$ ($m_3$) is the mass of the
neutrino state most (least) populated by the $\nu_e$ component.
Moreover, we only have an upper limit on the mixing angle
$\theta_{13}$, $\sin^2\theta_{13} \lsim 0.04 $, given by the 
CHOOZ reactor experiment~\cite{chooz}.

The supernova density profile is such that a neutrino oscillation
resonance in the $31$-channel, involving $\Delta m^2_{31}$ and
$\theta_{13}$, is bound to happen.  Whether this happens for neutrinos
or anti-neutrinos depends on the neutrino mass hierarchy (if normal or
inverted). Besides, the resonance is very sensitive to the parameter
$\theta_{13}$. The value of the hopping probability changes from zero
to one when $\theta_{13}$ goes from $10^{-4}$ to $10^{-1}$.  Clearly,
supernova neutrinos provide an excellent chance to determine two of
the neutrino unknowns: the mass hierarchy and $\theta_{13}$.  This
fact has been pointed out and explored by several 
authors~\cite{earthmatter,hiertheta13,sato}.

However, the open questions on the dynamics of the SN explosion could, 
in principle, plague the determination of neutrino properties. 
One might wonder if it will be possible to disentangle the 
uncertainties of the supernova physics from the uncertainties  
on the neutrino parameters, and  use experimental measurements of the 
SN neutrino fluxes to extract new information on both, the SN 
explosion mechanism and the neutrino oscillation parameters.

In this article we analyze in detail the prospects for extracting the
SN parameters as well as the neutrino oscillation parameters at three
different types of next-generation detectors, from the measurements of
neutrinos from the cooling phase of a galactic supernova. The most
realistic next-generation experiments under present consideration are
a megaton-scale water Cherenkov, a 100 kton liquid Argon and a 50 kton
scintillation detector. We will study the performance of each of these
detector types.  In our analysis we will vary a total of seven
parameters. Five are SN parameters: the average $\nu_e,\bar \nu_{e}$ and
$\nu_x$ energies, respectively, $\langle E_{e}\rangle$, $\langle
E_{\bar e}\rangle$ and $\langle E_{x}\rangle$; the ratio of 
the luminosities in $x$ and $e$ flavors, $\xi$ (we assume the 
$\bar \nu_e$ and $\nu_e$ luminosities to be equal) and finally 
the overall normalization of the fluxes, fixed by the total energy 
released ($E_b$) and the distance to the exploding star ($D$), $E_b/D^2$.  
The last two are neutrino oscillation parameters:
the value of the angle $\theta_{13}$ and the neutrino mass hierarchy.
We will simulate an observed set of data for given values of these
seven parameters and use a $\chi^2$ method to henceforth construct
confidence levels for the determination of these unknowns.  In
particular, we perform a comparison, highlighting strengths and
weaknesses of each type of proposed experiment. In this paper we only
consider the performance of the detectors regarding supernova
neutrinos. However, when making a decision of which detectors should
be build, their sensitivity to many other processes, including nucleon
decay, solar and atmospheric neutrinos, beta-beams and super-beams as
well as a neutrino factory, should naturally also be
analyzed~\cite{Rubbia:2004nf}.

There has been a number of earlier works on supernova neutrinos, with
many papers discussing only the extraction of neutrino parameters.
Refs.~\cite{barger,valle}, however, investigate the possibility to get
information on SN physics from the SN neutrinos using Super-Kamiokande
and SNO detectors. These analyses take into account fewer supernova
parameters and only some of the detection channels (considered as
inseparable) we will consider here. Also, the simultaneous analysis of
both normal and inverted hierarchy was not performed.  The analysis in
Ref.~\cite{Gil-Botella:2004bv} is very similar to the present study
but concerns only a liquid Argon experiment.  Bounds on neutrino
masses, from the delayed time-of-flight as a function of the neutrino
energy, obtainable with future large water Cherenkov as well as with a
liquid scintillation experiment has also been
discussed~\cite{Nardi:2004zg}.  Other methods to extract information
from a SN are: analyzing the earth matter effects
~\cite{Arafune:1987cj,earthmatter,Lunardini:2001pb,Dighe:2003vm,
  Mirizzi:2006xx,Guo:2006ap} which may occur if the supernova
neutrinos traverse the Earth (mantle/core) before reaching the
detector; studying the variation of particularly constructed
variables~\cite{Takahashi:2001ep,hiertheta13,Barger:2005it}, such as
ratios of average energies, and recently the possibility of observing
shock wave effects has attracted attention~\cite{Schirato:2002tg,
hiertheta13,shock,Fogli:2004ff,Fogli:2006xy,Choubey:2006aq}.

It should be noted that the determination of the SN parameters 
from the other studies, such as Earth matter effects are difficult, 
without prior knowledge of the value of $\theta_{13}$. This is due 
to the fact that the Earth matter effects dependent on a combination of the 
hopping probability and the difference in unoscillated neutrino fluxes. 
Therefore, apparently a full analysis varying both neutrino and 
SN parameters is necessary for obtaining information on the SN parameters.  
In the present paper we will show to which maximal accuracy the SN parameters
can be determined by the three types of experiments and how their
determination depends on the values of the unknown neutrino
parameters.  Such a determination will likely be very helpful for the
understanding of the physics of the core-collapse.  

In the next section we present the parameterized neutrino flux 
from the SN cooling phase. In Sec.~\ref{sec:method} we will discuss 
the analysis method and the three experimental setups we consider
here. In Sec.~\ref{sec:results} we discuss our results and we 
devote the final section to our conclusions.

\section{The neutrino flux from the cooling phase}
\label{sec:snflux}

In this analysis we will consider neutrinos emitted from the cooling
phase of a type II supernova.  This phase has the largest emission of
energy with an expected total (time integrated) luminosity, $E_b$, of
about $1-5 \times 10^{53}$ ergs. This luminosity is divided into all 6
flavors and we denote the individual contributions by $L_{i}$. We
assume (as usual) that $\nu_\mu$, $\nu_\tau$, $\bar \nu_\mu$, $\bar
\nu_\tau $ are indistinguishable$^{\dagger}$~\footnote{$^{\dagger}$
We neglect the small difference in $\nu_{i}$ and $\bar\nu_{i}$, 
$i=\mu,\tau$, fluxes originating from weak magnetism effects.} 
and denote them by the common index $x$.
Thus, we have $E_b = L_{e}+L_{\bar e} +4\, L_{x}$.  We will
furthermore assume that $\nu_e$ and $\bar \nu_e$ luminosities are
identical, {\it i.e.}, $L_e=L_{\bar e}$, which holds approximately in
most SN simulation.  We allow for a violation of luminosity
equipartition by defining the parameter
\begin{equation} 
      \xi= \frac{L_{x}}{L_{e}}\, .
\end{equation}
In general, simulations compute a value of $\xi$ between 0.5 and 
2~\cite{garching,livermore}. 

We use the pinched Fermi-Dirac distribution for the energy-spectra
of the neutrinos emitted from the supernova (the unoscillated flux)
\begin{equation}
   \phi_i^{0}(E) = \frac{1}{F_3(\eta_i) \, T_i^4}\,  
   \frac{E^2}{\exp (E/T_i-\eta_i) +1} 
   \;\;\;,   \quad i= e, \bar e, x
\label{fdflux}
\end{equation}
where $F_3$ is a normalization function.  The average energies,
$\langle E_i \rangle$, are linearly related to the temperatures once
the pinching parameter, $\eta_i$, is fixed.  As explained in the
introduction the supernova physics strongly suggests that these
average energies follow a hierarchy; $\langle E_{e} \rangle < \langle
E_{\bar e} \rangle < \langle E_{x} \rangle$.  The SN simulations
favor $0 \leq \eta_e,\eta_{\bar e} \leq 3$ and $\eta_x < 2$.  In our
simulations we take all $\eta$'s to be zero, corresponding to pure
Fermi-Dirac spectra. This is a conservative choice, since
superposition of narrower energy spectra will be easier for the
detectors to distinguish. In the case the average energies
vary with time, the value of $\eta$ will be smaller, thus justifying
our choice.  Naturally, it would be even better if one could vary also
the pinching parameters, but this is beyond our scope.

The unoscillated $\nu_i$ flux at distance $D$ from the supernova 
is given by 
\begin{equation} 
  F^0_{\nu_i} = \frac{L_i}{4\pi D^2} \,  \phi^0_{i} (E) \;,
\label{fluxatd}
\end{equation}
where all the luminosities are proportional to the total binding
energy $E_b$. It is worth remembering that one will normally not be 
able to see an optical counterpart to the SN (if the SN is inline with
the galactic center it will be obscured by dust), leaving the
distance $D$ unknown. A crude estimate of the chance that an optical
signal of the SN can be seen is only about one out of four. This
should be compared to the fact that less than four supernov{\ae} are
expected in our galaxy per century. Therefore, in this article we will
suppose that it is the combination $E_b/D^2$ that will be constrained
by the SN neutrino detection.  This can easily be translated to a
constraint on the total emitted energy if the distance $D$  can be
independently determined.  
Thus, the unoscillated flux of neutrinos from a supernova
is parametrized by 5 variables: $\langle E_{e} \rangle$, $\langle
E_{\bar e} \rangle$, $\langle E_{x} \rangle$, $\xi$ and $E_b/D^2$.

\begin{figure}[thpb]
\includegraphics[width=8.0cm,height=7.5cm]{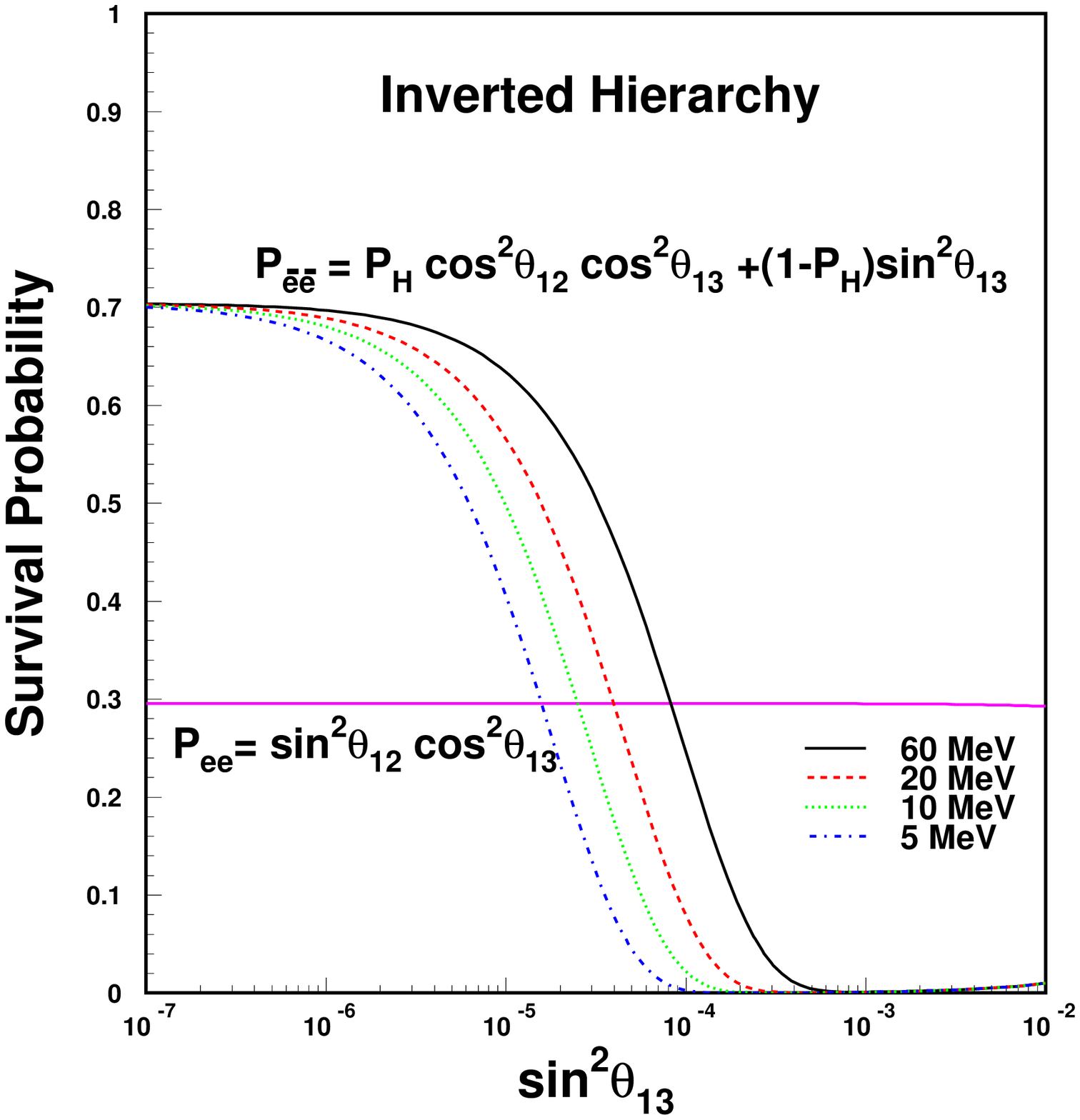}
\includegraphics[width=8.0cm,height=7.5cm]{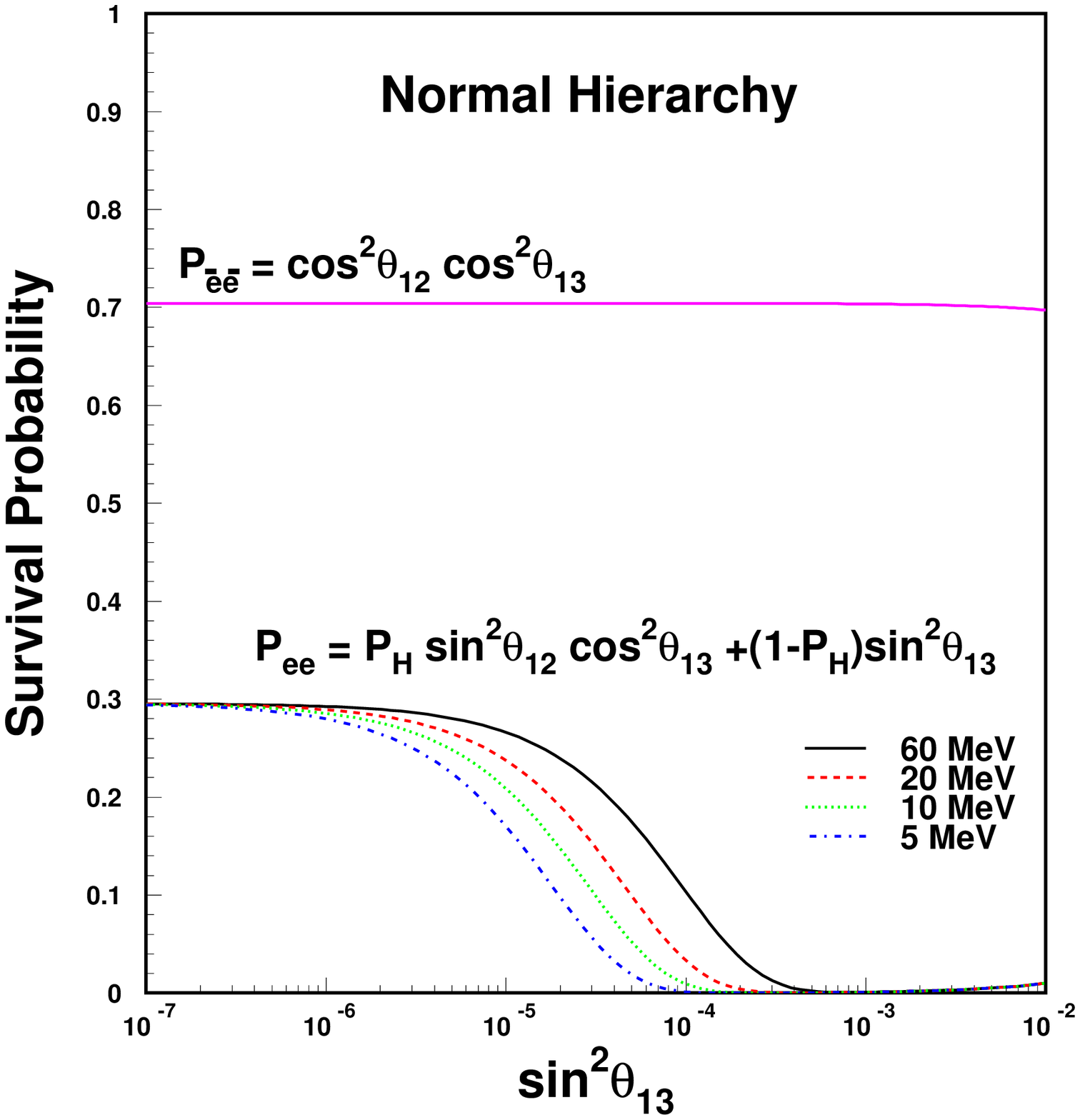}
\caption{The $\nu_e$, $P_{ee}$, and $\bar \nu_e$, $P_{\bar e \bar e}$,
  survival probabilities as a functions of $\sin^2 \theta_{13}$ for
  four different neutrino energies. We show the cases of inverted
  (left panel) and normal (right panel) hierarchy.  We have used
  $\sin^2\theta_{12}=0.3$ and $\vert \Delta m^2_{31}\vert=3.0 \times
  10^{-3}$ eV$^2$.}
\label{fig:SURP}
\end{figure}
The flux produced in the interior of the star will change its flavor
composition when traversing the outer parts of the star, due to
neutrino flavor oscillations in matter.  As we assume that the $\mu$
and $\tau$ fluxes are identical, the oscillated flux will only depend
on the $\nu_e$ and $\bar \nu_e$ survival probabilities.  These
probabilities strongly depend on the unknown neutrino parameters:
$\theta_{13}$ and the mass hierarchy.  We will now summarize these
effects assuming that the matter density of the SN scales as $\rho
\sim r^{-3}$, which seems to be the most realistic density
profile$^\S$\footnote{$^\S$For others density profiles see 
\cite{densprofile}.}.  We will use the approximation
\begin{equation}
  \rho(r) = C \cdot 10^{13} \left( \frac{10 \; \rm{km}}{r} \right) ^3 
\rm{g} \cdot \rm{cm}^{-3} \;\;\;\;, 
\label{density}
\end{equation}
and we take the value $C=4$.  Due to the mass gap ($\Delta
m^2_{21}/\vert \Delta m^2_{32}\vert \approx 1/30$) and the smallness
of $\theta_{13}$, the dynamics of the 3-$\nu$ system can be factorized
as two 2-$\nu$ sub-systems: a high (H) one, driven by $\Delta
m^2_{31}$ and $\theta_{13}$, and a low (L) one, driven by $\Delta
m^2_{21}$ and $\theta_{12}$. Correspondently, two resonances can
happen as neutrinos travel through the SN.  The H resonance will occur
for neutrinos (antineutrinos) in the case of normal (inverted)
hierarchy. In this case, there is a possible nonzero hopping 
probability $P_H$ for crossing between effective mass eigenstates. 
The L resonance will occur for
neutrinos$^\ddag$\footnote{$^\ddag$Since SNO's demonstration 
that the solar CC/NC ratio is less than $\frac{1}{2}$ we know 
that $\Delta m^2_{21}>0$.} and due to the values of the solar 
parameters given in (\ref{solarpar}) it will be always adiabatic 
(the level crossing probability $P_L \approx 0$). Thus, we will not 
discuss this resonance further.

The hopping probability $P_H$ can be parametrized as
\begin{equation}
  P_H = \exp \left[ -  \sin^2\theta_{13} 
  \left( \frac{1.08 \cdot 10^{7}}{E} \right)^{2/3}
  \left( \frac{|\Delta m_{31}^2|}{10^{-3}} \right)^{2/3}  C^{1/3} \right]
   \;,
\label{eq:ph}
\end{equation}
where $E$ is in MeV and $\Delta m_{31}^2$ in units of eV$^2$
~$^\dag$\footnote{$^\dag$Strictly speaking the $\Delta m^2$ in
  Eq.~(\ref{eq:ph}) should be $\Delta m^2_{32}$ for the normal
  hierarchy and $\Delta m^2_{31}$ for the inverted one. However since
  $\vert \Delta m^2_{31}\vert \approx \vert \Delta m^2_{32}\vert$, this
  is not relevant here.}.  
It should be noticed that the value of $C$ (which is uncertain by 
about a factor of 4) affects the translation from $P_H$ to
$\theta_{13}$.  Therefore, depending on how well the value of $C$ can
be determined, one might have to take this uncertainty into account
when putting bounds on $\theta_{13}$. Knowing the type of the
progenitor star would be a help in this
case.

The $\nu_e$ and $\bar \nu_e$ survival probabilities, $P_{ee}$ and 
$P_{\bar e \bar e}$, respectively, are approximated for the normal hierarchy by
\begin{eqnarray}
P_{ee}&\simeq & P_H |U_{e2}|^2 + (1 - P_H) |U_{e3}|^2 , \label{pue3nh} \\
P_{\bar e\bar e} & \simeq & |U_{e1}|^2  \label{pue3nh2}  \;,
\end{eqnarray}
where $U_{\alpha,i}\;,\alpha=e,\mu,\tau\; ,i=1,2,3$ are the 
Maki-Nakagawa-Sakata neutrino mixing matrix elements and we have used here 
the standard parameterization. Using (\ref{eq:param}) we find 
$|U_{e1}|^2\simeq 0.7$, $|U_{e2}|^2\simeq 0.3$ and $|U_{e3}|^2<10^{-2}$. 
And for the inverted hierarchy
\begin{eqnarray}
&&P_{ee} \simeq |U_{e2}|^2,
\label{p3ih2} \\
&&P_{\bar e\bar e} \simeq  P_H |U_{e1}|^2 + (1 - P_H) |U_{e3}|^2 \, .  
\label{p3ih}
\end{eqnarray}
These survival probabilities as a function of $\sin^2\theta_{13}$ 
are shown in Figure~\ref{fig:SURP} for various neutrino energies and 
the two types of hierarchies. We will neglect collective flavor 
transformation of the neutrinos, which in a detailed analysis should 
be added to the conventional MSW conversion in the supernova 
environment~\cite{Duan}.

The final fluxes arriving at Earth is simply given by 
\begin{eqnarray}
 F_{\nu_e} &=& F^{0}_{\nu_e} P_{ee} + F^{0}_{\nu_x} (1-P_{ee}), \\ 
 F_{\bar\nu_e} &=& F^0_{\bar\nu_e} P_{\bar e \bar e} 
 + F^0_{\bar \nu_x} (1-P_{\bar e\bar e}), \\
 F_{\nu_\mu} + F_{\nu_\tau} &=&  F^{0}_{\nu_e} (1-P_{ee}) 
+ F^{0}_{\nu_x} (1+P_{ee}), \\
 F_{\bar \nu_\mu}+ F_{\bar \nu_\tau} &=&
 F^0_{\bar \nu_e} (1- P_{\bar e \bar e}) 
 + F^0_{\bar \nu_x} (1+ P_{\bar e \bar e})\, .
\label{snfluxfinal}
\end{eqnarray}
Notice that in the case of luminosity equipartition and degeneracy 
in average energies, the flux is independent of the survival 
probabilities and thus also independent of the neutrino parameters. 

When calculating the oscillation probabilities we have neglected the
Earth matter effects~\cite{earthmatter}.  These effects can be
precisely calculated and thus will not affect our conclusions much, if
the direction of the supernova is known. Even in the case that an
optical counterpart of the SN can not be seen, the direction of the
supernova can be determined rather well from the neutrino flux, if the
detector (or another existing detector than the one being analyzed)
can measure the elastic scattering of neutrinos on electrons.  This
detection channel is highly forward peaked and in the case of
Hyper-Kamiokande the direction can be inferred to within $\sim
1^{\circ}$~\cite{Tomas:2003xn}. Therefore, we will not take into
account the Earth matter effects in our calculations.
However, in section \ref{sec:results} we will briefly discuss the 
possibility of extracting further information about the hierarchy 
by detecting Earth matter effects.

\section{\label{sec:method} The analysis method and the detectors}

In this section we will describe our method for studying the
sensitivity of the detectors to the parameters under investigation 
of neutrinos from a nearby supernova. While we are rather optimistic 
in our choice of detector performances we will be taking somewhat 
difficult choices for the input SN parameters.

As explained earlier, the neutrino fluxes arriving at the detectors
will depend on 7 parameters (5 SN dynamics parameters, and 2 neutrino
physics parameters).  For each experiment, we simulate the expected
number of neutrino events at each observable mode, for a fixed set of
these parameters. 
The artificially generated data will be our {\it input data} 
(the imagined true values).  We then
construct a $\chi^2=\chi^2(\langle E_{e} \rangle, \langle E_{\bar e}
\rangle, \langle E_{x} \rangle,\xi,
E_b/D^2,\sin^2\theta_{13},\text{sign}(\Delta m^2_{31}))$ function in
order to fit these unknown parameters to the {\it input data} in the
usual way. We only consider statistical uncertainties in our $\chi^2$.
This allows to compare the maximal attainable sensitivity for each detector 
type.
We compute the allowed regions for each pair of unknown parameters 
to estimate the experimental sensitivity to them by marginalizing 
with respect to the other 4 parameters, for a fixed hierarchy.
Since we construct the confidence level region for each hierarchy 
separately, our graphs will still be useful, in the case that the 
neutrino mass hierarchy is determined before a SN observation.
Moreover, this allows us to conclude whether or not the hierarchy can
be established.  

The artificially generated data are constructed simply by
calculating the theoretical expectation at the chosen set of input 
parameters. We have tested that this gives similar results 
as when a more realistic data set is used. We have constructed a 
Gaussian distributed data set, by choosing randomly a point 
from a Gaussian distribution centered around the expectation 
for the bin and of width equal to its  square-root. 
There is no difference in the results from the 
Gaussian distributed data set and the data set given by the
theoretical expectation, besides that the Gaussian data set 
might give a fluctuation of the allowed regions away from the 
central value. Generating the observed data set by the theoretical
expectation allows for easy independent reconstruction and 
comparison and we therefore prefer to 
use this simpler method for generating the data.

We will in this analysis use the
following parameters space when varying the SN parameters:
\begin{eqnarray}
&& \langle E_{e} \rangle \in [9-15] \;\; \rm{MeV} \;,\;\;\; 
\langle E_{\bar e} \rangle \in [12-17] \;\; \rm{MeV}  \;,\;\;\; 
\langle E_{x} \rangle \in [15-30] \;\; \rm{MeV} \;,  \nonumber \\
&& \xi  \in [0.5-2.0] \;,\;\;\; 
E_b/D^2 \in (2.0-4.0) \times 10^{51} \;\; \rm{ergs/kpc}^2 \;.  
\label{snparamarea}
\end{eqnarray}
This parameter space is roughly what is expected from SN 
simulations~\cite{garching,livermore}.  
In any case if an analysis of SN neutrino observation cannot 
determine the parameters within the area suggested in 
(\ref{snparamarea}), the SN simulations will probably do a better job. 
Naturally, the confidence levels and the accuracy with which 
each parameter can be measured, will depend on the {\it input data}. 
For instance, in the case when $\langle E_{e} \rangle =
\langle E_{\bar e} \rangle = \langle E_{x} \rangle$  and $\xi=1$ 
the neutrino flux becomes independent of the survival probabilities 
and thus of the neutrino oscillation parameters. 
Therefore, in this special limit (luckily 
not favored from SN simulations) the neutrino parameters cannot be 
deduced from the observation of SN neutrinos.  
We will discuss how the neutrino parameter determination depend 
on the strength of the hierarchy of the average energies.  
\begin{table}[htbp]
\centering
\begin{tabular}{|c||c|c|c|c|c|} \hline
 & $\langle E_{e} \rangle$ (MeV) & 
$\langle E_{\bar e} \rangle$  (MeV) & $\langle E_{x} \rangle$  (MeV)
& $\xi$ & $E_b/D^2$ (ergs/kpc$^2$) \\ \hline \hline
point 1 
& 12 &  15 &  18.0 & 1.50 & 3.0 $\times 10^{51}$  \\\hline
point 2 
& 12 &  15 &  18.0 & 0.75 & 3.0 $\times 10^{51}$ \\\hline
point 3 
& 12 &  15 &  16.5 & 1.50 & 3.0 $\times 10^{51}$ \\\hline
\end{tabular} 
\caption{Definition of the input SN parameters at some 
reference points.}   
\label{tab:points}
\end{table}

In Table \ref{tab:points} we define three points in the 
SN parameter space, that we will use as reference points. 
These points are situated within the expectation of the 
SN simulation given in equation \ref{snparamarea}. Most 
parameters are chosen at their central value. However,  
the value of $\xi$ has been chosen at two point slightly 
away from its central value (being 1), so as to see its 
impact on the accuracy with which the SN and neutrino 
parameters can be determined. Also the values 
of $\langle E_x \rangle$ is chosen at the lower range of its 
expectations. This has been done in order to see until which 
lower value a meaningful determination of the neutrino 
parameters can be expected.  
It is important to notice that increasing the value of 
$\langle E_x \rangle$ will increase the sensitivity of the 
detectors to the neutrino parameters as well as to the supernova 
parameters. First of all a higher value of $\langle E_x \rangle$ 
will give a larger number of events, since the detection 
cross-sections increases with energy. Also the increased 
hierarchy between $\langle E_x \rangle$ and $\langle E_{\bar e} \rangle$ 
will make the impact of the value of the survival probability larger.
For points 1 and 2 we study a rather conservative case of 
a hierarchy of only 19\% difference between 
$\langle E_{\bar e} \rangle$ and $\langle E_{x} \rangle$. 
For point 3 the hierarchy is of only 10\%, in which case it 
will become harder to determine the neutrino parameters. 
Although, such a weak hierarchy between the average energies seems 
rather unlikely it cannot yet be excluded.

The solar neutrino mixing angle and the atmospheric 
mass-square difference are fixed to values in the allowed 
region given in (\ref{eq:param}):
\begin{equation}
\theta_{12}=0.575 \; {\rm rad}  \;,\;\;\;\;  
|\Delta m^2_{31}|= 3.0 \times 10^{-3} \; {\rm eV}^2 \;.
\end{equation}
The atmospheric mixing angle does not enter into the calculation 
as $\nu_\mu$ and $\nu_\tau$ enter on the same footing both in 
the SN and in the detector (it introduces a unobservable rotation). 
Furthermore, as we do not take into account Earth matter effects, 
we have no dependence on the solar mass-square difference. 
When calculating the confidence levels we will just study the 
extreme cases for the true value of $\theta_{13}$. 
Meaning that we will only investigate two values of 
$P_H$; zero and one. In this case we have three scenarios for 
the input neutrino parameters:
\begin{enumerate}
\item Scenario i0: Inverted hierarchy and $P_H \simeq 0$ (large
  $\theta_{13}$);
\item Scenario n0: 
Normal hierarchy and $P_H\simeq 0$ (large $\theta_{13}$);  
\item Scenario a1: Any hierarchy and $P_H \simeq 1$ (small $\theta_{13}$).
\end{enumerate}
In the case of $P_H \simeq 1$ the inverted and normal hierarchy 
are identical. In the above scenarios large $\theta_{13}$ means 
a values corresponding to $\sin^2\theta_{13}=10^{-3}$ and small 
means a value corresponding to  $\sin^2\theta_{13}=10^{-6}$ 
(see figure~\ref{fig:SURP}).

We will analyze three different types of next-generation 
experiments, namely:
\begin{itemize}
\item Water Cherenkov (WaterC);
\item Liquid Argon (LAr);
\item Scintillation.
\end{itemize}
Each of these will be discussed in detail in the following 
subsections. The WaterC and Scintillation experiments are much more 
sensitive to anti-neutrinos than neutrinos, due to the dominant 
inverse beta decay detection. The dominant detection channel for 
a LAr detector on the other hand is charged current 
$\nu_e$ interactions on Argon, making it more sensitive to 
neutrinos.

Before discussing the details of each detector we will note 
some common features. First of all, obviously having sensitivity 
to more than one combination of neutrino fluxes will be essential 
to pin down the SN and neutrino parameters. 
For each experiment we will make contours for two scenarios: 
a pessimistic one, assuming that only the 
dominant channel can be used, and an optimistic one, where we assume 
several channels can be separated by the detector. This will illustrate 
the necessity of having sensitivities to several channels with 
different sensitivity to $\nu_e$/$\bar\nu_e$ and $\nu_x$ fluxes. 
In particular, a NC channel along with a CC channel will 
complement each other well. 
A true NC channel is independent of the neutrino  
parameters, as is easily seen, as we have  
\begin{eqnarray}
N_{\rm NC} & \propto & \int (F_{\nu_e} \sigma_{NC}+ F_{\bar\nu_e} \sigma_{NC}
+2F_{\nu_x} \sigma_{NC}+2F_{\bar\nu_x} \sigma_{NC}) \; dE\, , \\
& \propto & \int (F^0_{\nu_e}+ F^0_{\bar\nu_e}+4F^0_{\nu_x}) \sigma_{NC} \; dE\, ,
\end{eqnarray}
which is given purely in terms of the original fluxes. However, it 
should be remembered that eg. elastic scattering on electrons is 
a combination of NC and CC interactions for the electron neutrino 
and anti-neutrino species. Thus for this channel 
$\sigma_{\bar\nu_e} \neq \sigma_{\nu_e} \neq \sigma_{\nu_x}$ 
and the number of events have a slight variation with the 
oscillation parameters. 

We neglect any time-dependence of the neutrino fluxes and 
just look at the energy spectrum, or in the case that the 
energy cannot be measured, the total number of events integrated 
over time. The time-dependence of the energy spectra can  
be monitored by the experiments we study. Therefore, it will be possible to
get a feeling on how well the time-independence assumption works. 
The SN simulations suggest that a mild steady increase in the average 
energies as a function of time will occur. This suggest that indeed a 
pinched Fermi-Dirac spectrum, can still be used, but with a broader 
spectrum and thus a smaller value of $\eta_i$. Also a steady decrease 
in the luminosities is expected, not influencing our method. 

We will also assume that all detection efficiencies are 100\% above 
the threshold. This is a very optimistic assumption, but the efficiencies 
for these future experiments are presently unknown, and at least in this 
manner we treat all three experiments on the same footing. 
Also, we will neglect the energy resolution, but this is partly 
compensated by the use of wide bins in energy. We will use energy 
bins of 10 MeV, unless otherwise specified. 
The exact energy resolution, which is also unknown, can become
important in some cases where there are degeneracies between certain
parameters. However, for most of the parameter space the exact energy
resolution is not very important and will not change the general
result$^\sharp$\footnote{$^\sharp$ The energy resolution is, however, 
important for the determination of Earth matter effects for 
supernova neutrinos from a single detector}.

\subsection{Analysis of a Water Cherenkov Detector}
The water Cherenkov detectors have proven very successful, 
with the Super-K collaboration being first at announcing extremely 
compelling evidence of atmospheric neutrino oscillation  in 1998. 
The Hyper-Kamiokande~\cite{HyperK} (Hyper-K) detector is being 
proposed to replace the current Super-Kamiokande experiment and will 
have a total mass of about one megaton. 
Other proposed and more or less identical detectors, with the only
difference being their location and the exact mass, are the 
(American) UNO detector~\cite{Uno} and the (European) MEMPHYS 
detector~\cite{Memphys}. 
All these detectors will of course have similar sensitivities. 
We assume the fiducial volume of the WaterC to be 540 kton, which is 
the expectation for the Hyper-Kamiokande~\cite{HyperK} detector. 
Earlier works on the subject can be found in~\cite{barger,valle}
although these references only take into account one detection channel. 
As we will prove, the possibility to measure neutral current and 
charged current on oxygen as well as elastic scattering on electrons, 
will greatly improve the sensitivity of a WaterC detector.  

\begin{figure}[t]
\centering
\includegraphics[width=12cm,height=9cm]{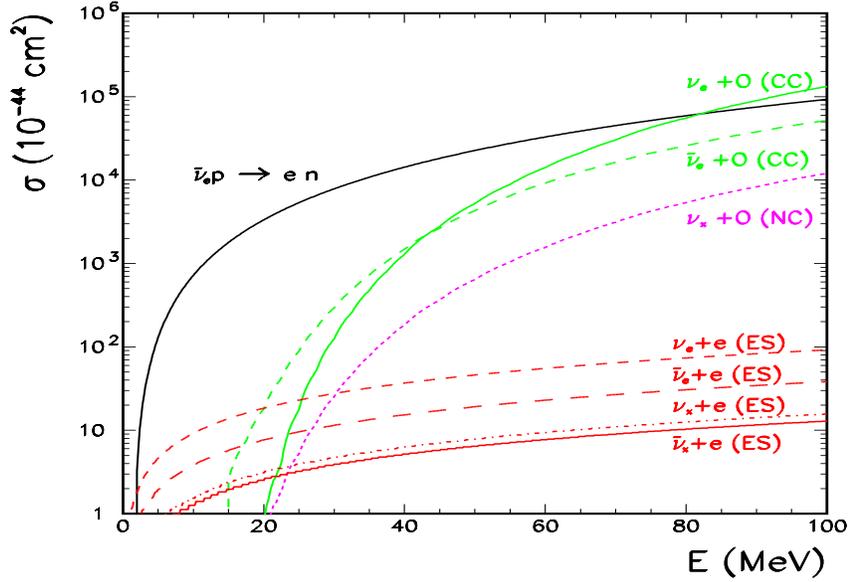}
\vspace{-0.8cm}
\caption{The cross-section for the various channels for 
neutrino detection in a WaterC detector. Here ES stands 
for elastic scattering on electrons.}
\label{figcrhk}
\end{figure}
We will take into account four different channels. 
Two channels are CC reactions that will provide spectral information 
for $\nu_e$ and $\bar\nu_{e}$ fluxes. One channel is a NC reaction and 
thus sensitive to all neutrino flavors, providing information on the 
total neutrino flux. The ELAS channel is, as discussed earlier, 
also mainly an NC reaction, but with a small CC contributions. 
Below we list the four channels.
\begin{enumerate}
\item
The inverse beta decay (IB) for detection of $\bar \nu_e$
\begin{equation}
  \bar{\nu}_e + p \rightarrow e^+ + n\,.
\end{equation}
This is by far the dominant detection channel and we will assume the 
threshold to be 5 MeV. The cross-section is well-known and we use the 
calculation given in Ref.\cite{vogel_anup}.
\item
The absorption of $\nu_e$ and $\bar \nu_e$ on oxygen 
by CC interactions (CC-O)
\begin{equation}
 \stackrel{\scriptscriptstyle (-)}{\nu}_e +^{16}{\rm O} \rightarrow 
 e^{\pm} + X \;.
\end{equation}
The cross-sections are taken from Ref.\cite{Kolbe:2002gk} 
and the threshold for detection is about 15 MeV. 
\item
The elastic scattering (ELAS) on electrons 
\begin{equation}
  {\nu}_i +e \rightarrow \nu_i + e\,\;\; ,
\end{equation}
is possible for all types of neutrinos, although the cross-sections
for $\nu_e$ and $\bar \nu_e$ are slightly higher due to the additional
CC contribution. The ELAS channel is easily separated 
as these events are strongly forward peaked. We have set the 
detection threshold at 7 MeV.
\item
An interesting channel for observation is the excitation of oxygen 
by NC interactions, followed by a decay chain with emission 
of a detectable mono-energetic photon, 
as first discussed in Ref.\cite{Langanke:1995he}, 
\begin{equation}
 \nu_i +^{16}{\rm O} \rightarrow 
 \nu_i + \gamma +  X \; .
\end{equation}
Hyper-K can detect photons with an energy greater than about 5 MeV. 
The excited $^{16}$O atom decay to either $^{15}{\rm O}^*$ or 
$^{15}{\rm N}^*$, which then emits a photon with the 
energy in the range 5-10 MeV. As the photon has a well defined 
energy, these events can be easily separated from the other 
detection channels.  We will refer to this channel as NC-O, 
even though it should be remember that it only includes those 
partial decay chains which give rise to a detectable 
mono-energetic photon. This channel differs from the others as the 
energy spectrum cannot be measured. Henceforth we only use the 
total number of events.  
\end{enumerate}

We will for the main parts of this article assume that 
the four channels can be separated, and make remarks 
on the case where only the inverse beta decay channel is detectable. 
The NC-O and ELAS channels should be easily separated from the
others. As discussed in Ref.\cite{Fogli:2004ff}, the IB events 
should be slightly forward peaked, whereas the CC-O events 
should be slightly backward peaked and this gives an opportunity 
to distinguish these events. Moreover, by addition of small amounts 
of gadolinium~\cite{Beacom:2003nk} in the WaterC detector, 
the capture of neutrons is possible and this would assure the 
separation of the IB and CC-O channels. This is also the reason that
we have put the detection threshold for the IB channel as low as 
5 MeV. Presumably even a lower threshold can be achieved with a 
gadolinium enriched WaterC detector.  
For the IB, CC-O and ELAS channels we calculate the
energy spectrum and we use 10 bins with a width of 
10 MeV and the first bin starting at 5 MeV. 
In Figure \ref{figcrhk} we show the cross-sections involved.

\begin{table}[tbp]
\centering
\begin{tabular}{|c||r|r|r|} \hline
\multicolumn{4}{|c|}{\bf Expected Number of Events in a 540 kton water 
Cherenkov detector} \\
\multicolumn{4}{|c|}{$\langle E_{e}
\rangle$ = 12 MeV, $\langle E_{\bar e} \rangle$ = 15 MeV, $\langle
E_{x} \rangle$ = 18 MeV, $L_x=1.5 \, L_e$}  \\ \hline
 & Any hierarchy & Inverted hierarchy  & Normal hierarchy \\
Reaction  & 
\multicolumn{1}{|c|}{Small $\theta_{13}$ } & 
\multicolumn{1}{|c|}{Large $\theta_{13}$} & 
\multicolumn{1}{|c|}{Large $\theta_{13}$}  \\ \hline \hline
{\bf Inverse beta decay} 
&  $1.4 \times 10^{5}$ & $2.1 \times 10^{5}$ & $1.4 \times 10^{5}$ \\ \hline 
{\bf $\stackrel{\scriptscriptstyle(-)}{\nu_e}$ CC on oxygen}  
&   $7.7 \times 10^{3}$  &  $10.7 \times 10^{3}$ & $9.5 \times 10^{3}$  \\ \hline  
{\bf ${\nu}_x +e \rightarrow \nu_x + e$} 
&   $8.4 \times 10^{3}$  &  $8.7 \times 10^{3}$ & $8.8 \times 10^{3}$  \\ \hline
{\bf NC on oxygen} 
&   $3.5 \times 10^{3}$ &  $3.5 \times 10^{3}$ & $3.5 \times 10^{3}$ \\\hline\hline
{\bf TOTAL} 
& $1.63 \times 10^{5}$ & $2.34 \times 10^{5}$  & $1.65 \times 10^{5}$ \\\hline
\end{tabular} 
\caption{Expected number of neutrino events in a 540 kton water Cherenkov 
detector for the SN parameters at point 1 and each of the cases a1, i0
and n0 for the neutrino parameters.
}   
\label{tab:rateshk}
\end{table}
In Table~\ref{tab:rateshk} we show the total number of events we
calculate for the SN parameters at point 1. The dominant channel
for Hyper-K is the inverse beta decay channel, which is only
sensitive to the $\bar \nu_e$ flux arriving at the Earth. However, the
NC-O, CC-O and ELAS channel is very important for having sensitivity 
to other combination of fluxes. Moreover, the ELAS channel gives the
opportunity to determine the direction to the supernova and thus an
early warning to astronomers will be possible. 
From Table~\ref{tab:rateshk} it is expected that if the true neutrino
parameters are consistent with the inverted hierarchy and large
$\theta_{13}$ then Hyper-K will be able to determine this scenario
with a very high confidence level.  But, even the sub-dominant channel
allows for a determination of the normal hierarchy if the angle
$\theta_{13}$ is sufficiently large. In section \ref{sec:results} we 
will comment on how this dependent on the SN parameters
(luminosities and average energies), by using our reference points.

\subsection{Analysis of a LAr Experiment}

Next, we will look at the possibility to determine the supernova
and neutrino parameters at a future Liquid Argon experiment. The are 
various LAr experiments proposed. The Icarus detector at CNGS is 
expected to have a 3 kton final version$^\dagger$~\footnote{
$^\dagger$Unfortunately, ICARUS might be interrupted due to 
cancellation of its funding.} 
and already a 300 ton detector is running. 
Moreover, the LANNDD ~\cite{LANNDD}, the
GLACIER~\cite{Glacier} and the Flare~\cite{Flare} detectors are being 
discussed as possible future detectors in the 100 kton size. 
Earlier works on the subject can be found in
\cite{Bueno:2003ei,Gil-Botella:2003sz,Gil-Botella:2004bv,Barger:2005it}.

\begin{figure}[ht]
\centering
\includegraphics[width=12cm,height=9cm]{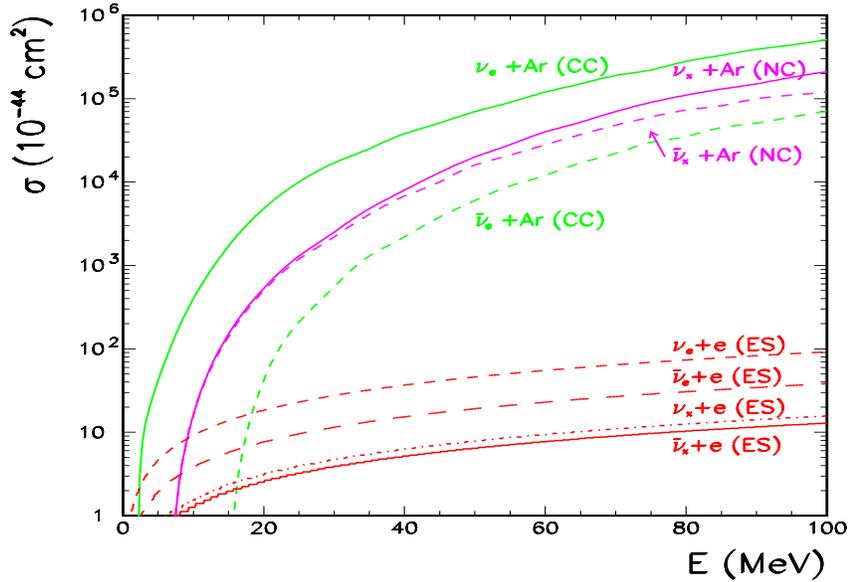}
\vspace{-0.8cm}
\caption{The cross-section for the various channels of 
neutrino detection in LAr.}
\label{fig:crla}
\end{figure}
We will take into account the following channels: 
\begin{enumerate}
\item 
Detection of $\nu_e$ through CC-interaction:
\begin{equation}
  \nu_e + ^{40}{\rm Ar} \rightarrow e^- + A^{\prime} + {\rm nN}
  \quad \quad (E_{\rm thr}=1.5 {\rm MeV}) \;,
\end{equation}
where ${\rm nN}$ represent emitted nucleons or other debris (like 
$\alpha$-particles etc.) and $A^\prime$ is the leftover nucleus. 
The CC cross-section on Argon are taken from 
Ref.~\cite{Kolbe:2003ys,Gil-Botella:2003sz}. 
\item
Detection of $\bar \nu_e$ through CC-interaction:
\begin{equation}
 \bar\nu_e +^{40}{\rm Ar} \rightarrow e^+ + A^{\prime} +{\rm nN}\, , 
\end{equation}
having a threshold about 7.5 MeV.
\item
The elastic scattering (ELAS) on electrons 
\begin{equation}
  {\nu}_x +e \rightarrow \nu_x + e\, ,
\end{equation}
is possible for all flavors of neutrinos. Again we have taken the
threshold to be 7 MeV.
\item
The scattering on Argon of any type of neutrino through NC-interaction:
\begin{equation}
 \nu_i +^{40}{\rm Ar} \rightarrow \nu_i +  A^{\prime} + {\rm nN}\, . 
\end{equation}
This channel has no sensitivity to the energy.  
The NC cross-section on Argon are taken from 
Ref.~\cite{Kolbe:2003ys,Gil-Botella:2003sz}.
\end{enumerate}

\begin{table}[tbp]
\centering
\begin{tabular}{|c||r|r|r|} \hline
 \multicolumn{4}{|c|}{\bf Expected Number of Events in a 100 kton LAr detector} \\
 \multicolumn{4}{|c|}{$\langle E_{e}
\rangle$ = 12 MeV, $\langle E_{\bar e} \rangle$ = 15 MeV, $\langle
E_{x} \rangle$ = 18 MeV, $L_x=1.5L_e$}\\ \hline
& Any hierarchy & Inverted hierarchy & Normal hierarchy  \\ 
Reaction   & \multicolumn{1}{|c|}{Small $\theta_{13}$} & 
\multicolumn{1}{c|}{Large $\theta_{13}$} & 
\multicolumn{1}{c|}{Large $\theta_{13}$} \\ \hline \hline
{\bf $\nu_e$ CC on Argon}      & $1.4 \times 10^{4}$ & $1.4 \times 10^{4}$
 & $1.7 \times 10^{4}$ \\ \hline 
{\bf $\bar \nu_e$ CC on Argon} &   $4.2 \times 10^{2}$ &   $7.9 \times
 10^{2}$ &   $4.2 \times 10^{2}$ \\ \hline
{\bf ELAS}                     &  $1.2 \times 10^{3}$ &  $1.3 \times
 10^{3}$ &  $1.3 \times 10^{2}$ \\\hline
{\bf NC on Argon}              & $1.3 \times 10^{4}$ & $1.3 \times 10^{4}$
 & $1.3 \times 10^{2}$ \\ \hline \hline
{\bf TOTAL}                    & $2.80 \times 10^{4}$ & $2.84 \times 10^{4}$
 & $3.16 \times 10^{4}$ \\\hline
\end{tabular}
\caption{Expected number of neutrino events at a 100 kton LAr detector 
for the SN parameters at point 1 and each of the cases a1, i0
and n0 for the neutrino parameters.
}   
\label{tab:ratesla}
\end{table}
As in Ref.\cite{Gil-Botella:2004bv} it is assumed that one can
separate all four channels. However, we will also make contours for
the `worse case' scenario were only detection in the $\nu_e$ CC
channel is available. For the $\nu_e$ and $\bar \nu_e$ CC reactions 
and the $\nu_i$ NC reactions, the energy and time-delay of the photons 
emitted  from the de-excitation of respectively K, Cl and Ar can be
used to classify the type of event. For the ELAS events no photons 
will be present. The cross-sections are shown in
Figure~\ref{fig:crla}, and it should be noted that we do not take into
account in our calculations their uncertainties. At the moment there
are no experimental confirmation of the theoretically calculated
cross-sections. But hopefully, in the case that a large scale LAr
detector will be realized, the cross-sections will already have been 
experimentally measured (eg. by ICARUS). 
For the channels with measurable energy spectra we again 
use 10 energy bins of 10 MeV each, the first bin starting at 5 MeV.  

In Table~\ref{tab:ratesla} we show the total
number of events for the SN parameters at point 1 for each
detection channel. The dominant channel is detection of $\nu_e$ 
by the charged current interaction on Argon. Also the NC and 
ELAS channels have a fairly large number of events, whereas the 
sensitivity to the $\bar \nu_e$ flux is rather weak. 

\subsection{Analysis of a Scintillation Detector}

Finally, we will examine the proposal for the 50~kton Low
Energy Neutrino Astronomy (LENA)~\cite{Ob05,LENA} 
liquid scintillation detector. 
We assume that LENA will be filled with pure PXE (C$_{16}$H$_{18}$). 
If another oil will be used the carbon to proton ratio may change 
and thus the results will change slightly. 
A discussion on SN neutrinos and 
scintillator detectors can be found in \cite{LENA,snscint}, 
although without an explicit calculation of the accuracy of the 
determination of the parameters.

\begin{figure}[htbp]
\begin{center}
\includegraphics[width=12cm,height=9cm]{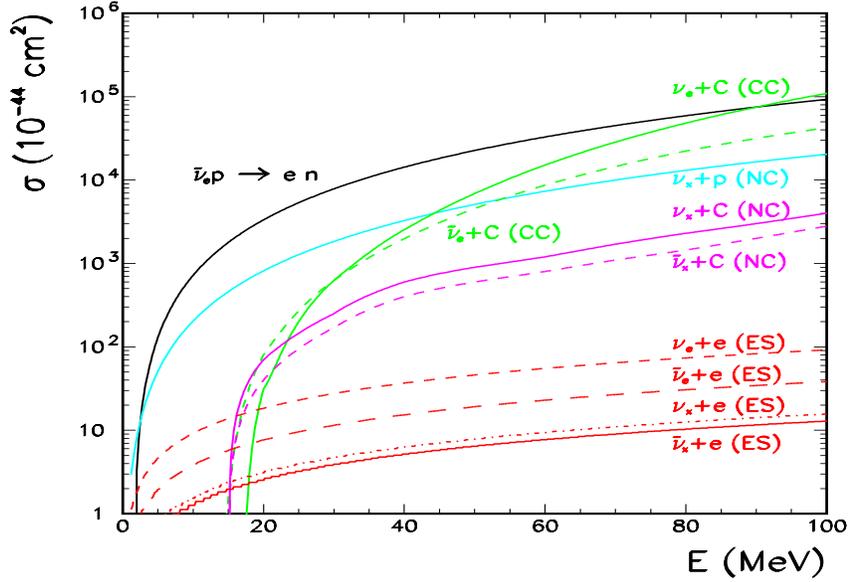}
\end{center}
\vspace{-0.8cm}
\caption{The cross-section for the various channels of neutrino 
 detection in LENA.}
\label{Fig:cross-section_lena}
\end{figure}
We will exploit six $\nu$ detection channels in LENA (three are CC 
reactions, two are NC reactions and the last is the ELAS channel),
these are listed below: 
\begin{enumerate}
\item The inverse beta decay for detection of $\bar \nu_e$
\begin{equation}
\bar{\nu}_e+p \rightarrow e^+ + n
\end{equation}
The threshold $\anue$ energy for this reaction
is 1.8 MeV.  Again we take the IB cross-section from \cite{vogel_anup}.
\item The CC capture of $\bar\nu_e$ on $^{12}$C
\begin{eqnarray}
&& \bar{\nu}_e + ^{12}{\rm C} \rightarrow \,  ^{12}{\rm B} + e^+ \;\;,\\
&& ^{12}{\rm B} \rightarrow ^{12}{\rm C}  + e^- + \bar{\nu}_e
\nonumber 
\end{eqnarray}
The threshold $\bar \nu_e$ energy for the capture on
$ ^{12}$C is 14.39 MeV. 
\item The CC capture of $\nu_e$ on $ ^{12}$C
\bea
&& \nu_e + ^{12}{\rm C} \rightarrow  ^{12}{\rm N} + e^- \;\;,\\
&& ^{12}{\rm N}  \rightarrow  ^{12}{\rm C} + e^+ + \nu_e
\eea
The threshold neutrino energy for capture on
$^{12}$C is 17.34 MeV.
\item Elastic scattering on protons 
\begin{equation}
\nu_i + p \rightarrow \nu_i + p
\end{equation}
This process might in some areas of parameters space even give 
a larger number of events than the IB process, due to the factor of 
six, originating from the number of neutrino and anti-neutrino species.
The cross-sections is taken from ~\cite{Beacom:2002hs}, where we 
have used the approximation of equal cross-section for neutrinos 
and anti-neutrinos. We have implemented a neutrino threshold 
energy of 25 MeV, corresponding roughly to a cut of 0.2 MeV 
in electron equivalent energy. 
\item NC scattering on $ ^{12}{\rm C}$:
\bea
&& \nu_i(\bar{\nu_i}) + ^{12}{\rm C} \rightarrow  ^{12}{\rm C^*}
+ \nu_i^\prime(\bar{\nu^\prime_i})  \;\;, \\
&& ^{12}{\rm C^*} \rightarrow  ^{12}{\rm C} + \gamma \;(15.11\; {\rm
  MeV}) \nonumber
\eea
The emission of a mono-energetic photon, makes this channel easily 
separated from the others.  
This cross-section is taken from \cite{vogel-12c}.
Since the emitted photon carries no information about the 
neutrino energy, this is the only channel for which LENA will 
have no energy information. Therefore, we will only use the total 
number of events from this process. 
\item
ELAS on electrons
\begin{equation}
{\nu}_{i}+e \to {\nu}_i +e \;,
\end{equation}
which has been discussed in earlier sections.
\end{enumerate}

\begin{table}[tbp]
\centering
\begin{tabular}{|c||r|r|r|} \hline
\multicolumn{4}{|c|}{\bf Expected Number of Events in a 50 kton Scintillation 
   Detector} \\
\multicolumn{4}{|c|}{$\langle E_{e}
\rangle$ = 12 MeV, $\langle E_{\bar e} \rangle$ = 15 MeV, $\langle
E_{x} \rangle$ = 18 MeV, $L_x=1.5 L_e$}  \\ \hline
 & Any hierarchy & Inverted hierarchy  & Normal hierarchy \\
Reaction  & 
\multicolumn{1}{|c|}{Small $\theta_{13}$} & 
\multicolumn{1}{|c|}{Large $\theta_{13}$} & 
\multicolumn{1}{|c|}{Large $\theta_{13}$} \\ \hline \hline
{\bf Inverse beta decay} 
&  $1.0 \times 10^{4}$ & $1.5 \times 10^{4}$ &  $1.0 \times 10^{4}$ \\ \hline 
{\bf ${\bar \nu_e}$ CC on carbon}  
&  $6.0 \times 10^{2}$ &  $1.1 \times 10^{3}$  &  $6.0 \times 10^{2}$  \\ \hline  
{\bf ${\nu_e}$ CC on carbon}  
&  $1.0 \times 10^{3}$  &  $1.0 \times 10^{3}$ &  $1.4 \times 10^{3}$ \\ \hline  
{\bf ${\nu_i}$ NC on proton}  
&  $9.9 \times 10^{3}$ & $9.9 \times 10^{3}$ & $9.9 \times 10^{3}$\\ \hline  
{\bf ${\nu}_i +e \rightarrow \nu_i + e$} 
&  $7.9 \times 10^{2}$ & $8.2 \times 10^{2}$ & $8.2 \times 10^{2}$ \\ \hline
{\bf ${\nu_i}$ NC on carbon} 
& $1.4 \times 10^{3}$ & $1.4 \times 10^{3}$ &  $1.4 \times 10^{3}$ \\ \hline\hline
{\bf TOTAL} 
& $2.39 \times 10^{4}$ & $2.93 \times 10^{4}$ & $2.43 \times 10^{4}$ \\\hline
\end{tabular} 
\caption{Expected number of neutrino events in a 50 kton scintillation  
detector for the SN parameters at point 1 and each of the cases a1, i0
and n0 for the neutrino parameters.
}   
\label{tab:rateslena}
\end{table}
Here again we take two approaches. The conservative one, 
where we consider only the inverse beta decay channel 
and the optimistic one,  where we assume that all channels 
can be distinguished from each other.
In principle, one can hope that, due to the distinctive 
signatures of the above discussed channels they can 
be separated. The most doubtful discrimination is between the 
$\nu_e$ and $\bar \nu_e$ CC reactions on carbon. It might be 
possible to separate these by using the delayed coincidence of the 
$\beta^+$/$\beta^-$ decays with the primarily produced
electron/positron and the knowledge of the average lifetimes of 
the produced unstable nuclei. 
For the channels with sensitivity to the neutrino energy spectra 
(all except the NC scattering on carbon) we again use 10 energy bins 
of 10 MeV each, the first bin starting at 5 MeV.  
Moreover, due to the very fine energy resolution expected for LENA,
with a threshold of order 200 keV, we also include an extra low-energy 
bin, with the events originating from neutrino energies below 5 MeV, 
for the IB and ELAS channels.

The dominant channels for LENA are the inverse beta 
decay channel (for $\bar\nu_e $) and  the $\nu_i$ NC scattering on
protons. In Table ~\ref{tab:rateslena} we show the 
total number of events for the SN parameters at 
point 1 for each observable neutrino channel in LENA. 
The detector will have a large sensitivity to both $\bar \nu_e$ 
and the total neutrino flux and even a reasonable sensitivity to
the $\nu_e$ flux. 

\section{Results}
\label{sec:results}
\begin{figure}[p]
\centering
\includegraphics[width=17cm,height=19cm]{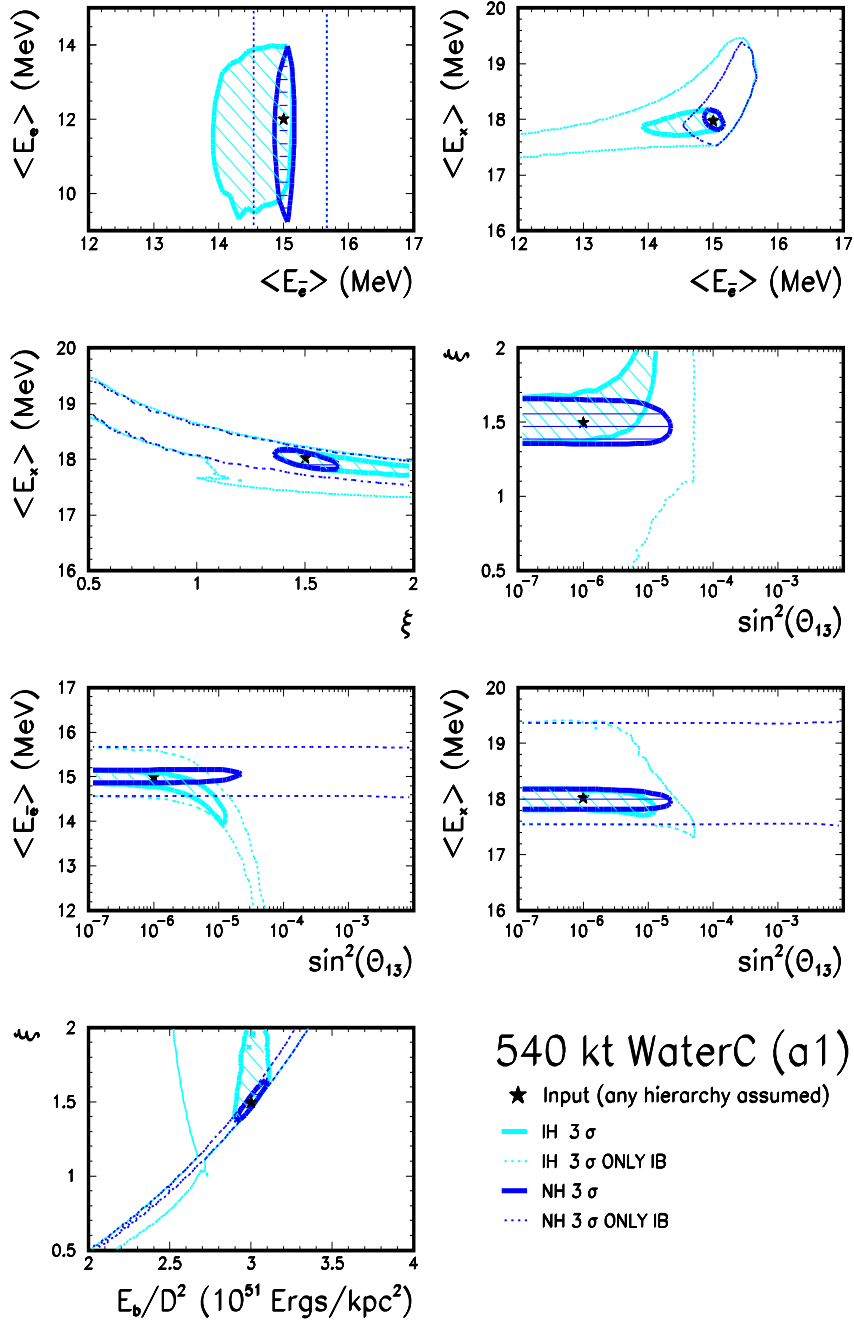}
\caption{Sensitivity of a 540 kton WaterC detector assuming true 
SN parameters as in point 1 of Table \ref{tab:points}, 
for any hierarchy and $\sin^2(\theta_{13})=10^{-6}$ ($P_H \simeq 1$).
We show  3$\sigma$ CL contours (2 dof) using:
all 4 channels (IB + CC-O + ELAS+ NC-O) and normal hierarchy (NH)
marked by the dark (blue) horizontally hatched area; 
only the IB channel and normal hierarchy 
marked by the dark (blue) dashed line; 
all 4 channels and inverted hierarchy (IH) 
marked by the light (cyan) diagonally hatched area; 
only the IB  channel and inverted hierarchy 
marked by the light (cyan) dashed line.}
\label{fig:hki1}
\end{figure}
\begin{figure}[p]
\centering
\includegraphics[width=17cm,height=19cm]{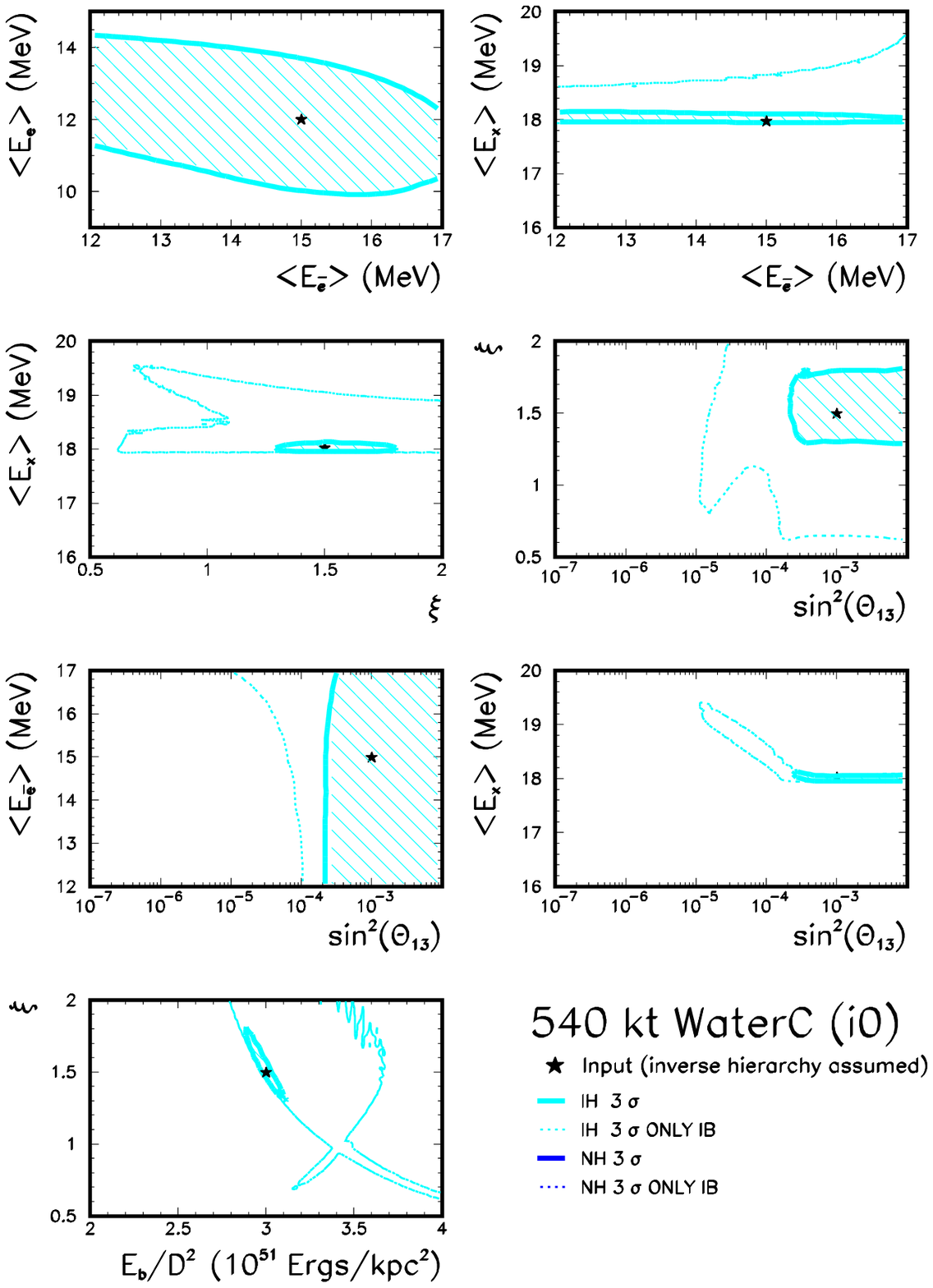}
\caption{Same as Fig.~\ref{fig:hki1} but for the
inverted hierarchy and  $\sin^2(\theta_{13})= 10^{-3}$ 
($P_H \simeq 0$). The normal hierarchy is ruled out by more 
than 5$\sigma$, having a global $\chi^2_{\rm min} \simeq 900$ 
in the case of all four channels being present.  
}
\label{fig:hki0}
\end{figure}
\begin{figure}[p]
\centering
\includegraphics[width=17cm,height=19cm]{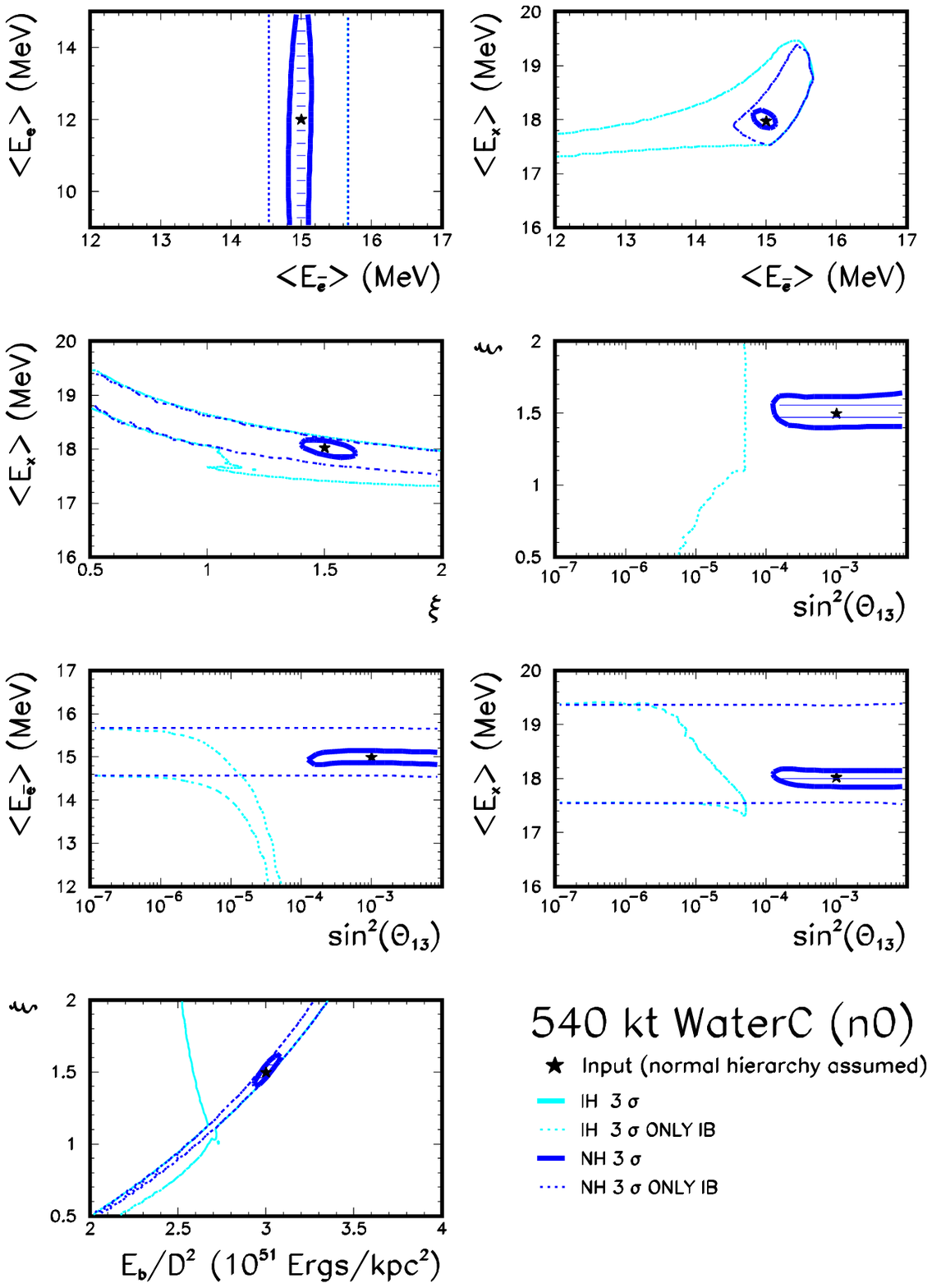}
\caption{Same as Fig.~\ref{fig:hki1} but for the 
normal hierarchy and $\sin^2(\theta_{13})=10^{-3}$ ($P_H \simeq 0$).
The inverted hierarchy is ruled out by more than 
5$\sigma$, having a global $\chi^2_{\rm min}=160$
in the case of all four channels being present.
} 
\label{fig:hkn0}
\end{figure}
\begin{figure}[p]
\centering
\includegraphics[width=17cm,height=19cm]{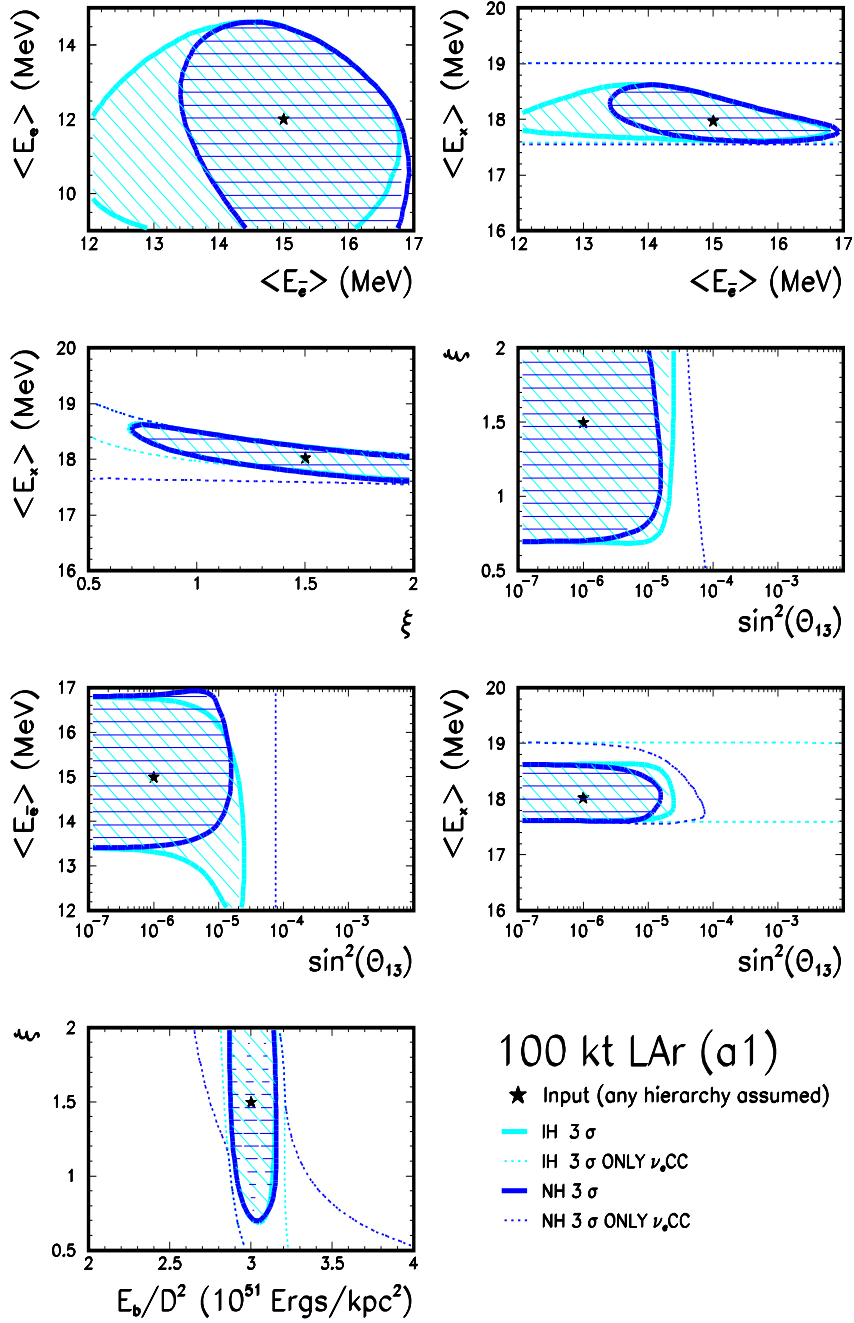}
\caption{Sensitivity of a 100 kton LAr detector assuming true 
SN parameters as in point 1 of Table \ref{tab:points}, for 
any hierarchy and $\sin^2(\theta_{13})=10^{-6}$ ($P_H \simeq 1$).
We show  3$\sigma$ CL contours (2 dof) using: all 4 channels 
($\nu_e$CC+ $\bar\nu_e$CC + ELAS + NC) and normal hierarchy 
marked by the dark (blue) horizontally hatched area; only the $\nu_e$CC
channel and normal hierarchy marked by the dark (blue) dashed line; 
all 4 channels and inverted hierarchy marked by the light (cyan) 
diagonally hatched area; only the $\nu_e$CC channel 
and inverted hierarchy marked by the light (cyan) dashed line.}
\label{fig:lai1}
\end{figure}
\begin{figure}[p]
\centering
\includegraphics[width=17cm,height=19cm]{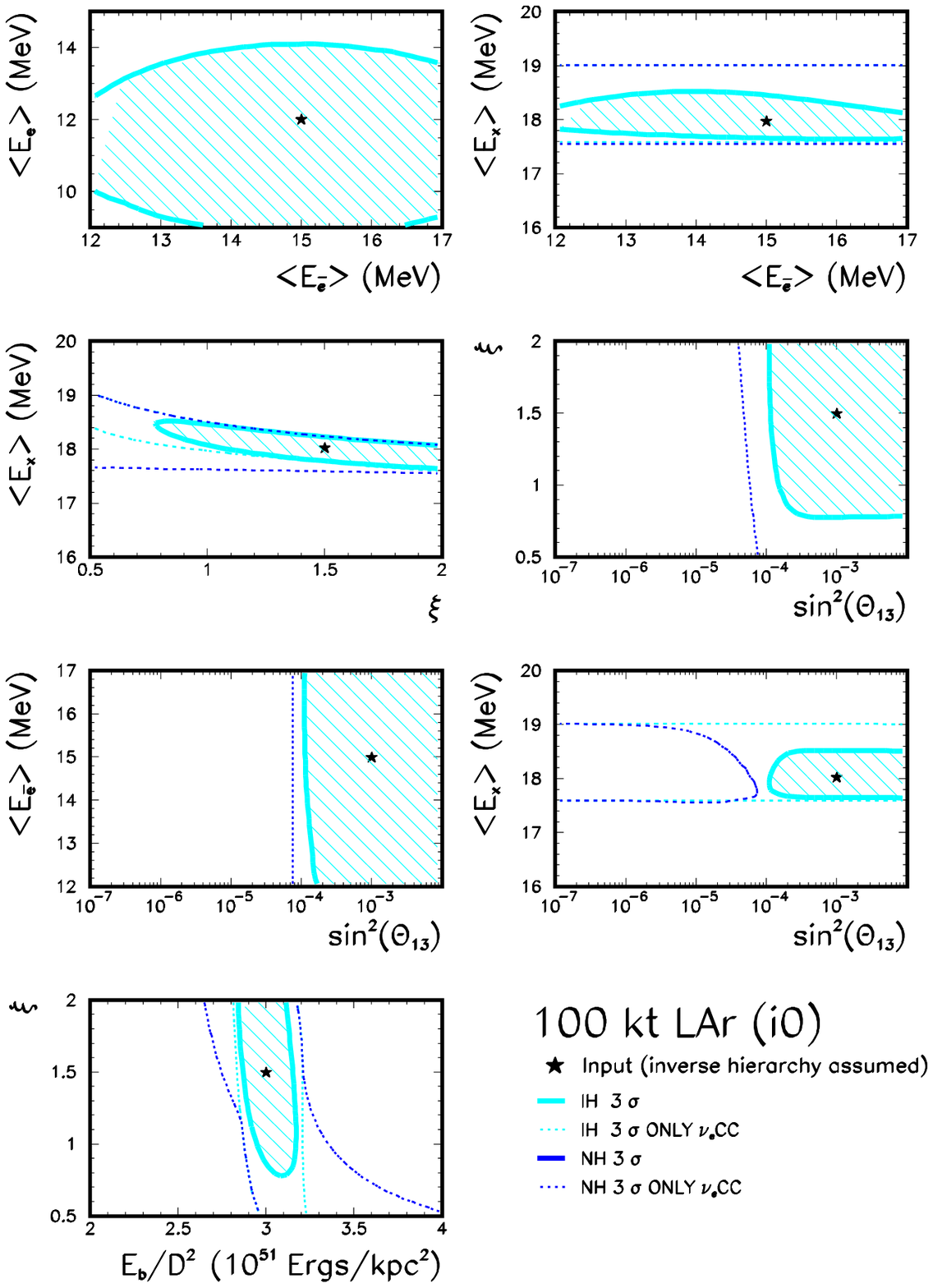}
\caption{Same as Fig.~\ref{fig:lai1} but for the
inverted hierarchy and $\sin^2(\theta_{13})=10^{-3}$ ($P_H\simeq 0$). 
The normal hierarchy is ruled out by more than 4$\sigma$,
having a global $\chi^2_{\rm min}=20$
in the case of all four channels being present.  
}
\label{fig:lai0}
\end{figure}
\begin{figure}[p]
\centering
\includegraphics[width=17cm,height=19cm]{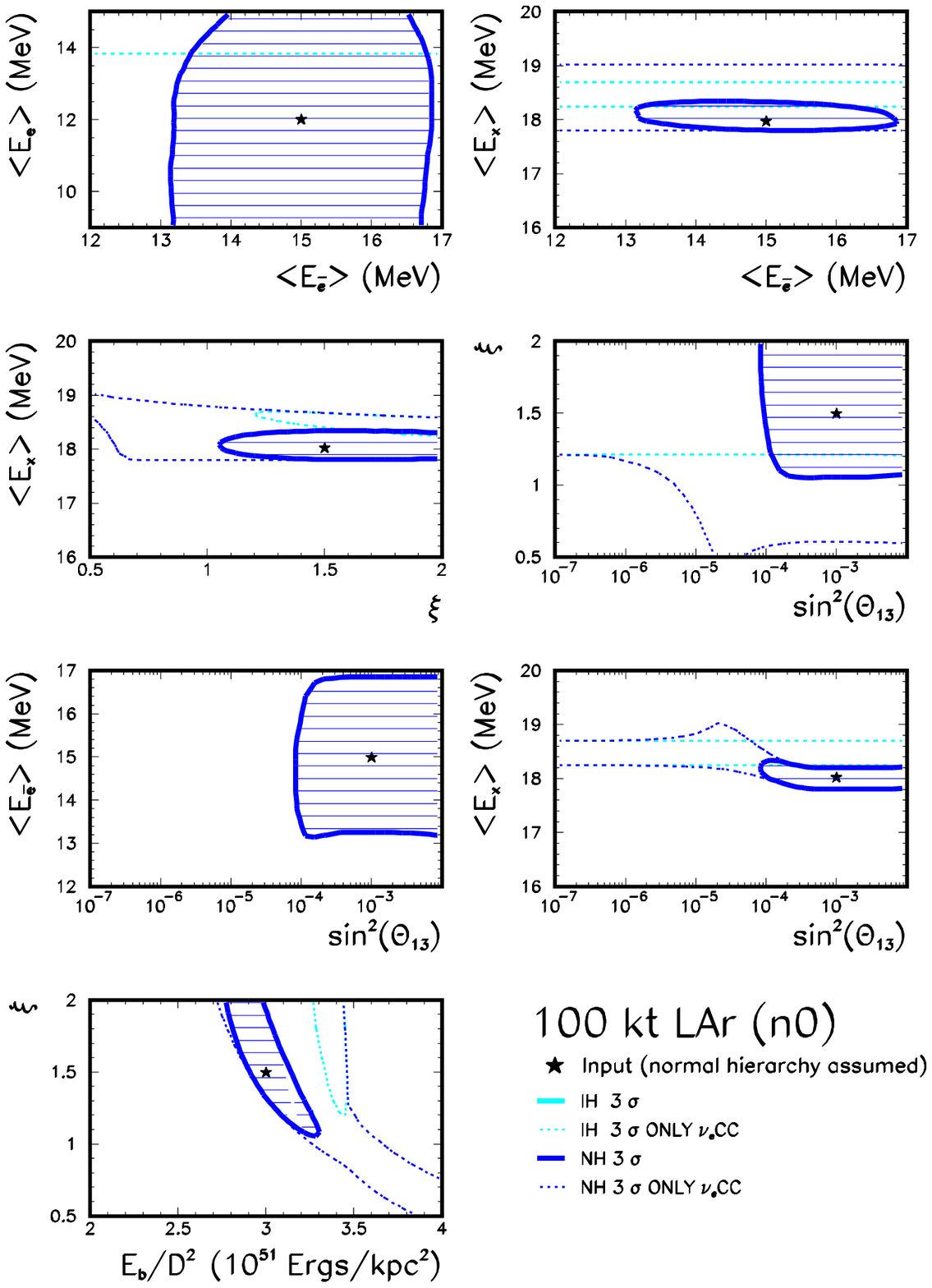}
\caption{Same as Fig.~\ref{fig:lai1} but for the 
normal hierarchy and $\sin^2(\theta_{13})=10^{-3}$ ($P_H\simeq 0$). 
The inverted hierarchy is ruled out by more than 5$\sigma$,
having a global $\chi^2_{\rm min}=280$
in the case of all four channels being present.  
}
\label{fig:lan0}
\end{figure}
\begin{figure}[p]
\centering
\includegraphics[width=17cm,height=19cm]{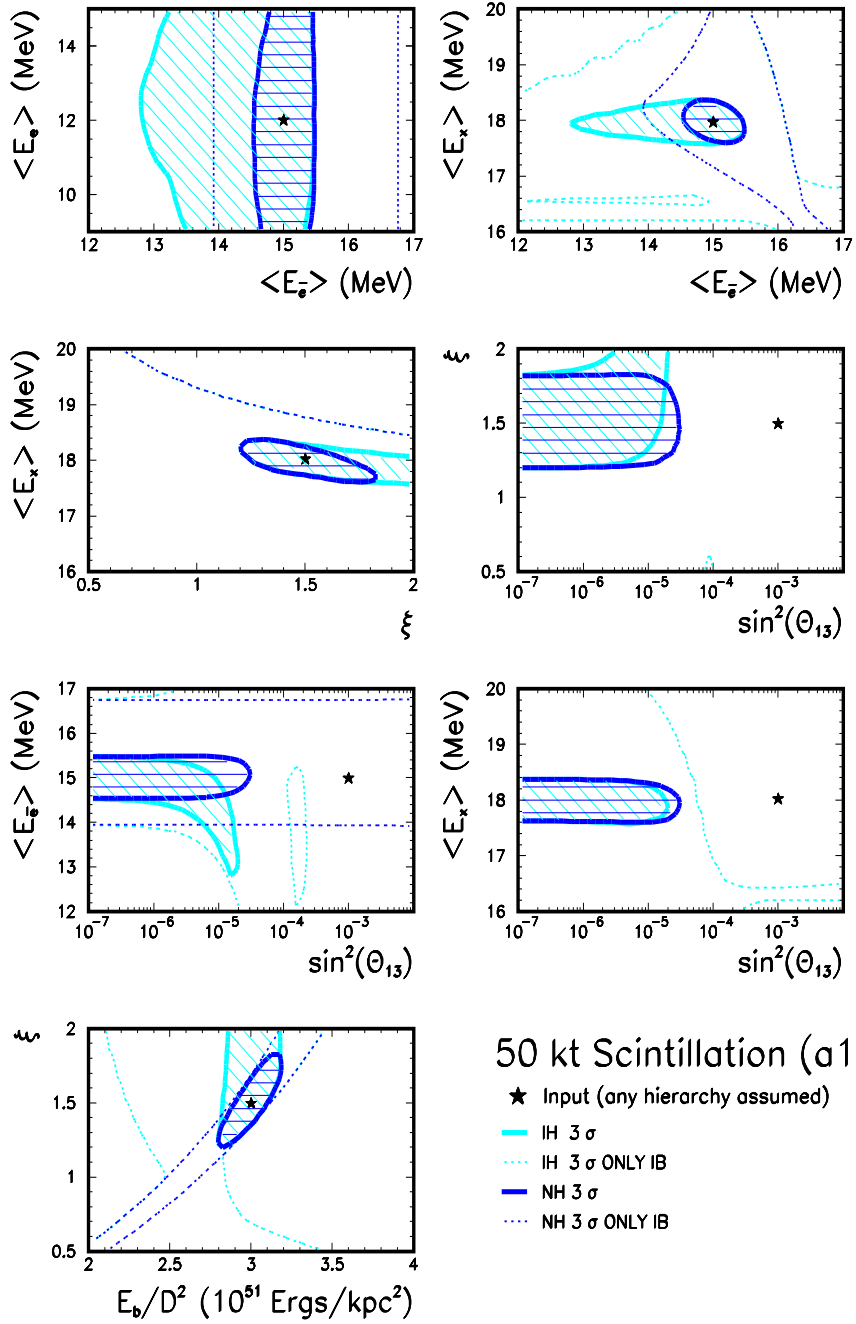}
\caption{Sensitivity of a 50 kton Scintillation detector assuming 
true SN parameters as in point 1 of Table \ref{tab:points}, for 
any hierarchy and  $\sin^2(\theta_{13})= 10^{-6}$  ($P_H \simeq 1$).
We show  3$\sigma$ CL contours (2 dof) using: all 6 channels 
(IB + $\nu_e$CC+ $\bar\nu_e$CC + $\nu-p$ + NC + ELAS) and normal hierarchy 
marked by the dark (blue) horizontally hatched area; only the IB
channel and normal hierarchy marked by the dark (blue) dashed line; 
all 4 channels and inverted hierarchy marked by the light 
diagonally (cyan) hatched area; only the IB channel 
and inverted hierarchy marked by the light (cyan) dashed line.
}  
\label{fig:sci1}
\end{figure}
\begin{figure}[p]
\centering
\includegraphics[width=17cm,height=19cm]{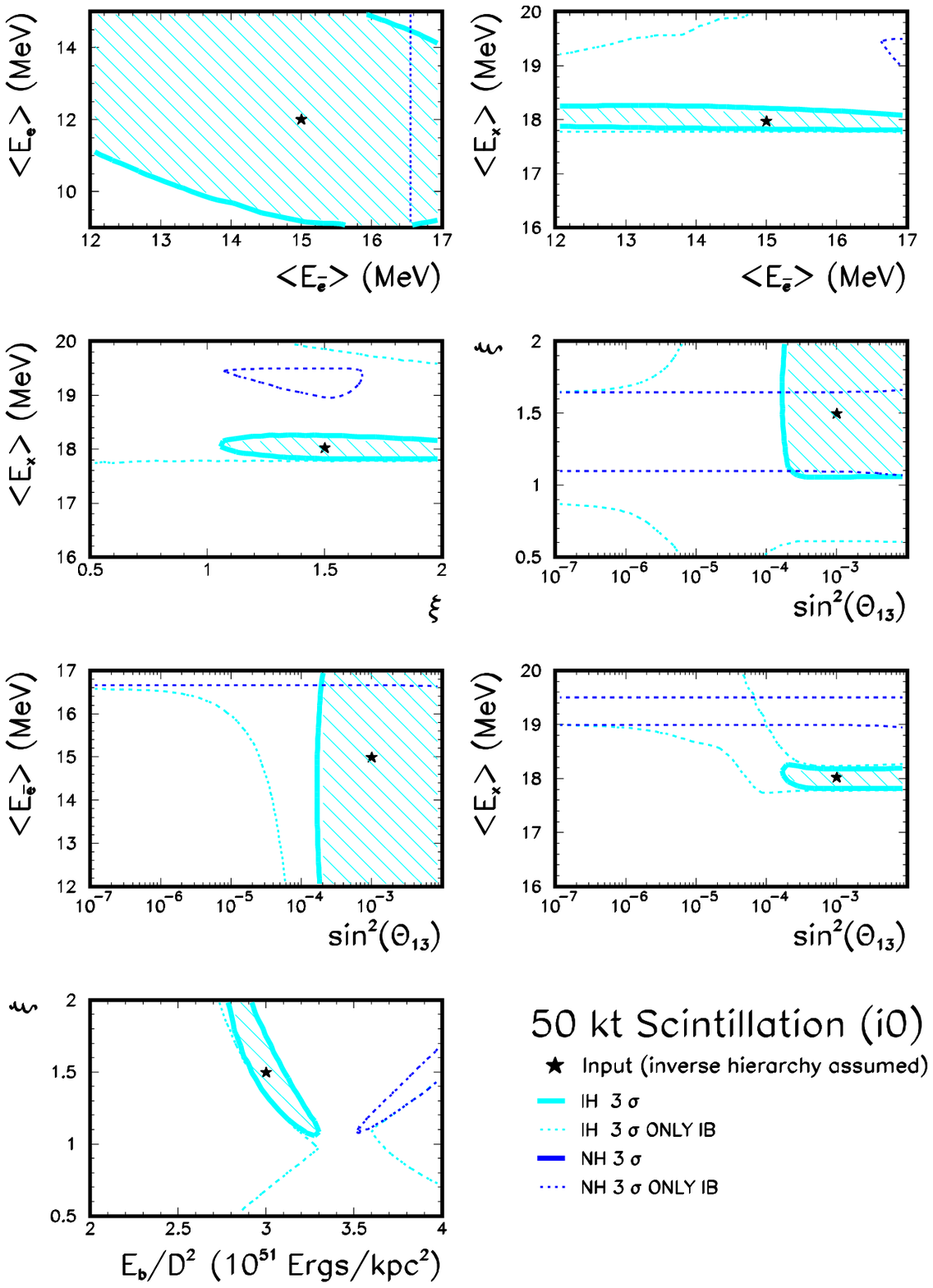}
\caption{Same as Fig.~\ref{fig:sci1} but for the 
inverted hierarchy and  $\sin^2(\theta_{13})= 10^{-3}$  ($P_H \simeq 0$).
The normal hierarchy is ruled out by more than 5$\sigma$,
having a  global $\chi^2_{\rm min}=120$
in the case of all six channels being present.}  
\label{fig:sci0}
\end{figure}
\begin{figure}[p]
\centering
\includegraphics[width=17cm,height=19cm]{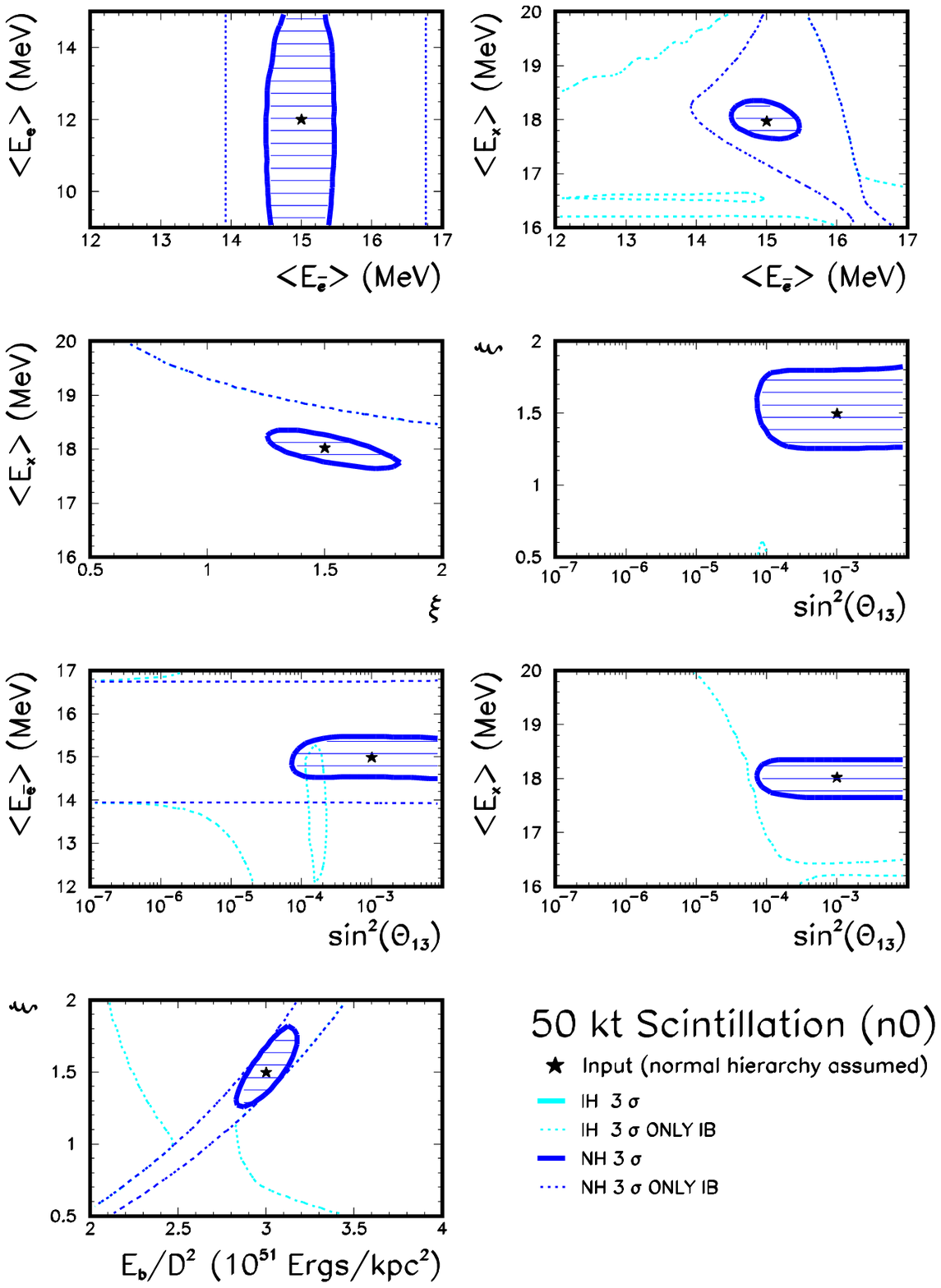}
\caption{Same as Fig.~\ref{fig:sci1} but for the 
normal hierarchy and  $\sin^2(\theta_{13})= 10^{-3}$  ($P_H \simeq 0$).
The inverted hierarchy is ruled out by more than 5$\sigma$,
having a global $\chi^2_{\rm min}=60$
in the case of all six channels being present.}  
\label{fig:scn0}
\end{figure}

In this section we will study two scenarios for the detection 
capability for each detector: the case where only the 
main CC channel is available and the case where all channels 
discussed in section \ref{sec:method} are available and 
moreover separable.  
As detecting only the main CC channel is a rather pessimistic scenario, 
we include in appendix A a discussion of the detector performance 
in a few other cases with a more realistic, but non-optimal, 
detector-setup. 
The main reasons for studying the main CC channel only case 
is to compare to other studies (Refs.\cite{barger,valle}) and also 
to illustrate the importance of having both NC and CC channels. 
Furthermore, this is the absolute worst case scenario and with 
the two contours considered in this section we give the span of 
possible allowed regions. 

In Figs.~\ref{fig:hki1}, \ref{fig:hki0} and \ref{fig:hkn0}
we show the 3$\sigma$ CL allowed regions obtained for a 
WaterC detector in the three limiting cases a1, i0 and n0, 
discussed in Sec.~\ref{sec:method}. 
The corresponding regions for the LAr and the Scintillation 
detectors are shown in respectively
Figs.~\ref{fig:lai1}, \ref{fig:lai0} and  \ref{fig:lan0} and 
Figs.~\ref{fig:sci1}, \ref{fig:sci0} and  \ref{fig:scn0}.

In the case a1 the hierarchy cannot be determined, as both hierarchies
produce the same neutrino fluxes.  All three experiments present a
good sensitivity to $\theta_{13}$ as can be seen from Figs.
\ref{fig:hki1}, \ref{fig:lai1} and \ref{fig:sci1} and will provide 
an upper limit on $\sin^2\theta_{13}$ of about 1-2 $\times 10^{-5}$, 
if there are several channels available. 
In the case that only the main channel is available (IB or
$\nu_e$CC) a degeneracy in $\theta_{13}$ and the hierarchy occurs and
$\theta_{13}$ cannot be determined (unless the hierarchy has already 
been established by another experiment).   
This independence of $\theta_{13}$ for the case of normal hierarchy
and the WaterC and Scintillator detectors, is evident from 
Fig.\ref{fig:SURP} as the IB channel is only sensitive to 
$P_{\bar e \bar e}$, which is almost constant 
as a function of $\theta_{13}$. Similarly, the LAr detector is not 
sensitive to $\theta_{13}$ in the case that only the $\nu_e$CC 
channel is available and the hierarchy is inverted. 
The anti-neutrino flux arriving at Earth for the case a1 (and n0) 
consist of roughly 70\% with a temperature of 15 MeV and 30\% 
with a temperature of 18 MeV (for point 1). 
For the Scintillator detector the statistics for the only IB 
case is such that a pure flux with temperature of about 16.5 MeV 
can also fit this data. Therefore, the inverted hierarchy is allowed 
even for large values of $\theta_{13}$. However, for a small interval 
around $\sin^2\theta_{13}\simeq 10^{-4}$ and low values of 
$\langle E_{\bar e} \rangle$ it is not possible to fit the energy 
spectrum, giving rise to islands in the contours as seen 
on Fig.\ref{fig:sci1} (the exactly same contours are found in 
Fig.\ref{fig:scn0}).  
If $\sin^2\theta_{13}$ turns out to be in the region where 
$P_H \simeq 1$, then there is no possibility
that it can be determined by any of the laboratory experiments
currently proposed~\cite{reactor_white,theta13,beta-beam,nufactory}.
In this case SN information will be extremely valuable.
If several channels can be included in the analysis, the 
value of the SN parameters $\langle E_{x}\rangle$, 
$\langle E_{\bar e}\rangle$ (except for the LAr detector) and 
$E_b/D^2$ can be determined to a few \% level in case a1. 
On the other hand, $\xi$ and $\langle E_{e} \rangle$ will be 
much less constrained by data (for point 1).

If $\sin^2\theta_{13}$ is large, {\it i.e.} in the region where 
$P_H \simeq 0$ (cases i0 and n0), all three experiments can determine 
the neutrino mass hierarchy. From Figure \ref{fig:hki0} and Figure
\ref{fig:hkn0} we clearly see that if $\theta_{13}$ is large, the
neutrino mass hierarchy can be determined with a high confidence level
independent of the true hierarchy, in a WaterC detector. Also the LAr
and the Scintillation detectors can establish the hierarchy in the
large $\theta_{13}$ region, as shown in Figs.~\ref{fig:lai0}-\ref{fig:lan0} 
and \ref{fig:sci0}-\ref{fig:scn0}, although with less significance.
It should be noted that, as we will discuss below, 
the establishment of the hierarchy depends on the parameter 
space assumed for the SN parameters, and in particular whether 
$\langle E_{\bar e} \rangle$ is allow to undertake the same 
value as $\langle E_{x} \rangle$. 

In the case i0, the WaterC detector can determine the hierarchy 
and will give a lower limit on $\sin^2\theta_{13}$, even if 
only the IB channel is available. 
For the LAr detector and the scenario with only the $\nu_e$CC 
channel being present, we see that the normal hierarchy can 
explain the i0 scenario. This is again evident from 
Fig.\ref{fig:SURP} as in this case the information 
contained in the measurement is that $P_{ee} \simeq 0.3$, which 
occur for the normal hierarchy and small values of $\theta_{13}$ 
and for inverted hierarchy and any value of $\theta_{13}$. 
For the only IB case and the Scintillator detector, the measurement 
contain the information of the $\bar \nu_e$ flux with a 
temperature corresponding to the value of $\langle E_x \rangle$. 
This can also be fitted with 
a normal hierarchy and raising the value of 
$\langle E_x \rangle$ (giving about 30\% of the flux) and 
$\langle E_{\bar e} \rangle$ (giving about 70\% of the flux) 
such as to simulate a pure $\bar \nu_e$ flux of 18MeV. 
The statistics of the WaterC detector is good enough to reject this 
case for the SN parameter space given in Eq.\ref{snparamarea}.
If all channels are considered in the data analysis,  
the LAr and Scintillation detectors can produce a similar constraint 
on $\sin^2\theta_{13}$ as the WaterC detector and the lower limit 
is given by 1-2 $\times 10^{-4}$. 
For scenario i0 the $\bar \nu_e$ flux arriving at the Earth is 
identical to the original $\nu_x$ flux and thus in this case 
one looses much of the sensitivity to $\langle E_{\bar e} \rangle$.  
But, a slightly better sensitivity to $\langle E_{x} \rangle$ 
is obtained as compared to the a1 case, where a part of 
$F_{\bar \nu_e}$ is the original $\bar \nu_e$ flux. 
The LAr detector can measure  $\langle E_{x} \rangle$ (even if 
only $\nu_e$CC events are used) and $E_b/D^2$ to about 5 \%, but 
will not be very sensitive to the other SN parameters.
Also, the LENA-type detector, can determine $\langle E_{x} \rangle$ 
and $E_b/D^2$ to a few \% with the help of all channels.
The WaterC detector will be very sensitive to all SN parameters, 
except for $\langle E_{\bar e}\rangle$, if all channels contribute.

In the case n0, even if only the main channels are available, 
$\langle E_{x} \rangle$ can be determined to
a few \% by the WaterC detector (see Fig.\ref{fig:hkn0}). 
In this scenario there is basically no sensitivity to
$\langle E_{e} \rangle$, as the $\nu_e$ flux is identical 
to the original $\nu_x$ flux. But, all the others SN parameters 
can be accessed with very good precision by the WaterC detector, 
if all channels take part in the analysis. The LAr setup can 
measure $\langle E_{x} \rangle$ to a few \% and 
$\langle E_{\bar e} \rangle$ and $E_b/D^2$ to about 10 \%
(see Fig.\ref{fig:lan0}). A LENA-type detector will have a
sensitivity of about 5 \% for $\langle E_{x} \rangle$, 
$\langle E_{\bar e} \rangle$ and $E_b/D^2$ and about of 20\% 
for $\xi$.
Furthermore, if all channels are considered in the data analysis,  
the LAr and Scintillation detectors can again provide similar 
bounds on $\sin^2\theta_{13}$ as the WaterC detector, giving 
a lower limit of 0.7-2 $\times 10^{-4}$.  
If only the IB channel is available not even the WaterC detector can 
determine the hierarchy. As seen from Fig.\ref{fig:SURP} the value 
of $P_{\bar e \bar e}$ of roughly 0.7 is also found for the inverted 
hierarchy and small values of $\theta_{13}$.

Let us shortly compare the sensitivity to the $\theta_{13}$ angle 
to that of other proposed future experiments. 
One expects the reactor experiment Double Chooz to reach down to 
$\sin^2\theta_{13} \simeq 5\times 10^{-3}$, and the next generation 
$\theta_{13}$ reactor experiment Daya-Bay to reach 
$\sin^2\theta_{13} \simeq 2.5\times 10^{-3}$~\cite{reactor_white}. 
The proposed novel technique which exploits the recoilless resonant 
absorption of $\bar \nu_e$ to measure $\sin^2 \theta_{13}$ in a 
short baseline experiment, may be able to reach similar 
sensitivity~\cite{theta13}. So to reach sensitivities 
to $\sin^2\theta_{13} \lsim 10^{-3}$ before a new galactic 
SN observation one would probably need beta-beams~\cite{beta-beam} 
or neutrino factories~\cite{nufactory}. Correspondingly, the 
determination of a lower bound on $\theta_{13}$ of order $10^{-4}$ 
from a SN observation for the cases i0 and n0, can be of 
great importance.

In the following we shortly compare to the work in 
Ref.\cite{valle}, where only the inverse beta decay channel at a WaterC 
detector was analyzed. In this paper only the case a1 is studied 
and the $\theta_{13}$ angle is not varied. This seems unlikely as 
there are no known experiment that can restrict $\theta_{13}$ to be 
smaller than $10^{-6}$, which would be necessary for getting 
independence of the exact value of the CHOOZ angle. Therefore, it is 
erroneously concluded in Ref.\cite{valle} that the value of $\xi$ can 
be well determined at HyperK with only the IB channel available. 
As can be seen from Fig.\ref{fig:hki1} there is a degeneracy between 
$\theta_{13}$ and $\xi$, only broken by the addition of NC channels. 
This indetermination of $\xi$ in turn influence the measurement of 
$E_b$, as there is a degeneracy between $E_b$ and $\xi$ (see 
Fig.\ref{fig:hki1} panel 9). Therefore, overall the allowed regions 
found in Ref.\cite{valle} are too restrictive as compared to 
our contours for the case of only IB.  

Next we compare to the work in Ref.\cite{barger}, 
where the WaterC detector with the inverse beta channel 
along with the CC-O was studied, although these were 
assumed inseparable. In addition the SNO detector was 
also analyzed, but as the fiducial mass is much smaller, 
this experiment does not add significantly to the statistics. 
The main sensitivity in Ref.\cite{barger} is to the 
$\bar \nu_e$ flux with only a minor sensitivity to the $\nu_e$ 
flux. In this paper it is concluded that the HyperK detector 
cannot determine the normal hierarchy even for large 
$\theta_{13}$. Clearly, we don't agree with this statement, as
with the addition of the sub-dominant channels this is indeed 
possible. However, we see that if only the IB channel is 
available this conclusion is true. But, it must be remembered 
that the only-IB scenario is not a realistic one. We also note 
that for small $\theta_{13}$ the very precise determination 
of $\langle E_{\bar e} \rangle$ claimed in Ref.\cite{barger} 
is only possible for the 
case that normal hierarchy has already been established.

\begin{table}[tbp]
\centering
\begin{tabular}{|c||ccc|ccc||} \hline
& \multicolumn{3}{c|}{point 2} & \multicolumn{3}{c||}{point 3} \\ 
& i0 & n0 & a1 & i0 & n0 & a1  \\ \hline  \hline
$\langle E_e \rangle$ IH & 11.0-13.2 & -- & 11.1-13.4 
& 10.0-14.1 & -- & $<14.2$ \\ 
$\langle E_e \rangle$ NH & -- & 10.8-13.9 & 10.5-13.0 
& $<13.8$   & NO & $<14.0$  \\ \hline
$\langle E_{\bar e} \rangle$ IH & 12.6-16.2 & -- & 14.5-15.2
& NO        & -- & 14.4-15.2 \\ 
$\langle E_{\bar e} \rangle$ NH & -- & 14.9-15.2 & 14.9-15.2 
& 16.4-16.8 & 14.7-15.2 & 14.8-15.2 \\ \hline
$\langle E_x \rangle$ IH & 17.9-18.1 & -- &  17.8-18.2
& 16.4-16.6 & -- &  16.2-16.7 \\ 
$\langle E_x \rangle$ NH & -- & 17.8-18.1 &  17.7-18.1
& 16.0-16.8 & 17.3-17.7 & 16.3-16.7 \\ \hline
$\xi$ IH & 0.7-0.8 & -- & 0.6-0.8 
& 1.3-1.7 & --       & $> 1.3 $ \\ 
$\xi$ NH & -- & 0.7-0.8 & 0.7-0.8 
& 0.7-0.9 & 1.4-1.6 & 1.3-1.7 \\ \hline
$E_b/D^2$ IH & 2.9-3.1  & -- &  2.9-3.1
& 2.8-3.2 & -- & 2.8-3.2 \\ 
$E_b/D^2$ NH & -- & 2.9-3.1 &  2.9-3.1
& 2.9-3.2 & 2.9-3.1 & 2.8-3.2 \\ \hline
$\sin^2\theta_{13}$ IH & $>2 \cdot 10^{-4}$ & -- & $<1\cdot 10^{-5}$  
& $>1\cdot 10^{-4}$  & -- & $<2\cdot 10^{-5}$ \\ 
$\sin^2\theta_{13}$ NH & -- & $>2\cdot 10^{-4}$ & $<2\cdot 10^{-5}$ 
&  $>1\cdot 10^{-5}$ &  $>8\cdot 10^{-5}$  & $<3\cdot 10^{-5}$ \\ \hline
Hier. det. & yes & yes & no & no & yes & no \\ \hline
Hier. excl. & $>5\sigma$ & $>5\sigma$ & -- & -- & $>5\sigma$ & --  \\ \hline
\end{tabular} 
\caption{The allowed parameter space at 3$\sigma$ CL (2 dof) for the 
points 2 and 3 of  Table~\ref{tab:points} by a 540 kt  
WaterC detector using all channels. The symbol `--' is used when 
there is no allowed area. In this case, we have written the 
confidence level with which a given hierarchy can be excluded. 
The symbol `NO' is used when there are no restrictions on the 
parameter space given in Eq.~(\ref{snparamarea}).
The energies are in MeV and $E_b/D^2$ in units of 
10$^{51}$ ergs/kpc$^2$.}   
\label{tab:WAp23}
\end{table}

\begin{table}[tbp]
\centering
\begin{tabular}{|c||ccc|ccc||} \hline
& \multicolumn{3}{c|}{point 2} & \multicolumn{3}{c||}{point 3} \\ 
& i0 & n0 & a1 & i0 & n0 & a1  \\ \hline  \hline 
$\langle E_e \rangle$ IH & 10.0-13.4 & -- &  $ <13.8$
& $< 14.1$ & --         & $<14.2$  \\ 
$\langle E_e \rangle$ NH & --         & $< 14.6$ &  $ <13.8$
& $< 12.7$ & NO        & $<14.2$ \\ \hline
$\langle E_{\bar e} \rangle$ IH & $>$ 13.2 & -- &  13.3-16.4
&   NO     & --         & 12.2-16.8 \\ 
$\langle E_{\bar e} \rangle$ NH & --  & 13.8-16.2 & 14.0-16.4 
& $>15.7$  & 13.2-16.8 & 13.4-16.8 \\ \hline
$\langle E_x \rangle$ IH & 17.4-18.6 & -- &  17.3-18.6 
& 16.1-17.0 & --        &  16.2-17.2 \\ 
$\langle E_x \rangle$ NH & -- & 17.8-18.5 &  17.4-18.6
& 16.2-16.7 & 16.3-16.9 & 16.2-17.0 \\ \hline
$\xi$ IH & $<$ 1.4  & -- &  $< 1.5$
& $>0.7$    & --         & NO  \\ 
$\xi$ NH & --  & 0.6-1.1 &  $< 1.3$
& 0.7-1.3   & $>1.0$    & $>0.7$ \\ \hline
$E_b/D^2$ IH & 2.8-3.2  & -- &  2.8-3.2
& 2.8-3.2   & --    & 2.8-3.1 \\ 
$E_b/D^2$ NH & -- & 2.7-3.3 &  2.8-3.3
& 2.9-3.4   & 2.7-3.3   & 2.8-3.2 \\ \hline
$\sin^2\theta_{13}$ IH & $> 6 \cdot 10^{-5}$  & -- & $< 5 \cdot 10^{-5}$ 
& $>6\cdot 10^{-5}$ & --   & $< 2\cdot 10^{-5}$ \\ 
$\sin^2\theta_{13}$ NH & -- & $> 6 \cdot 10^{-5}$ &  $< 2 \cdot 10^{-5}$ 
& 1-8$\cdot 10^{-5}$ & $>6\cdot 10^{-5}$ & $<4 \cdot 10^{-5}$ \\ \hline
Hier. det. & yes & yes & no & no & yes & no \\ \hline
Hier. excl. & $>5\sigma$ & $3 \sigma$ & -- & -- & $>5\sigma$ & --  \\ \hline
\end{tabular} 
\caption{Same as Table~\ref{tab:WAp23} but for a 100 kt   
LAr detector.}
\label{tab:LAp23}
\end{table}

\begin{table}[tbp]
\centering
\begin{tabular}{|c||ccc|ccc||} \hline
& \multicolumn{3}{c|}{point 2} & \multicolumn{3}{c||}{point 3} \\ 
& i0 & n0 & a1 & i0 & n0 & a1  \\ \hline  \hline 
$\langle E_e \rangle$ IH & 10.1-14.2 & -- &  $<14.8$
& NO & --    & NO  \\ 
$\langle E_e \rangle$ NH & --         & $<14.3$ &  $<14.7$
& NO & NO   & NO \\ \hline
$\langle E_{\bar e} \rangle$ IH & 12.2-16.8  & -- & 13.6-15.3 
& NO & --         & 13.6-15.6 \\ 
$\langle E_{\bar e} \rangle$ NH & --  & 14.6-15.3 & 14.6-15.3 
& NO & 14.5-15.5 & 14.5-15.5 \\ \hline
$\langle E_x \rangle$ IH & 17.7-18.3  & -- &  17.5-18.7
& 16.1-17.0  & --        & 16.0-16.9  \\ 
$\langle E_x \rangle$ NH & -- & 17.6-18.4 &  17.4-18.4
& 15.9-16.9  & 16.2-16.8 & 16.1-16.9 \\ \hline
$\xi$ IH & 0.6-1.0   & -- &  $<0.8$
& $<0.8$ {\it or} $>1.0$  & --    & $>1.1$  \\ 
$\xi$ NH & --  & 0.6-0.9 &  $<0.9$
& 0.6-1.0   & 1.2-1.9   & 1.1-1.9 \\ \hline
$E_b/D^2$ IH & 2.7-3.3  & -- & 2.8-3.3
& 2.8-3.3  & --    & 2.8-3.2 \\ 
$E_b/D^2$ NH & -- & 2.8-3.2 & 2.8-3.2$> 10^{-4}$
& 2.8-3.3  & 2.8-3.2  & 2.8-3.2 \\ \hline
$\sin^2\theta_{13}$ IH & $> 1 \cdot 10^{-4}$  & -- & $< 2\cdot 10^{-5}$ 
& $< 10^{-5}$ {\it or} $> 10^{-4}$   & --   & $< 2\cdot 10^{-5}$ \\ 
$\sin^2\theta_{13}$ NH & -- & $> 6 \cdot 10^{-5}$ &  $< 4\cdot 10^{-5}$ 
& NO & $>5\cdot 10^{-5}$ & $< 5\cdot 10^{-5}$ \\ \hline
Hier. det. & yes & yes & no & no & yes & no \\ \hline
Hier. excl. & $>5\sigma$ & $>5\sigma$ & -- & -- & $>5\sigma$ & --  \\ \hline
\end{tabular} 
\caption{Same as Table~\ref{tab:WAp23} but for a 50 kt  
Scintillation detector.}
\label{tab:SCp23}
\end{table}

In Tables \ref{tab:WAp23}, \ref{tab:LAp23} and \ref{tab:SCp23} 
we show the allowed parameter space for the input point 2 of 
Table~\ref{tab:points}. For point 2, we have chosen a 
smaller value of $\xi$, making the original $\nu_e$ and 
$\bar \nu_e$ fluxes twice as large as the original $\nu_x$ 
flux. There is not much difference between the results of 
point 1 and 2, except that the determination of 
$\langle E_{\bar e} \rangle$ and $\langle E_{e} \rangle$ 
are better for point 2. The accuracy of the determination of 
$\langle E_{x} \rangle$ has little dependence on the value of 
$\xi$, although it increases slowly for larger values of $\xi$.

In Tables \ref{tab:WAp23}, \ref{tab:LAp23} and \ref{tab:SCp23} 
we show the results for point 3 of Table \ref{tab:points}. 
For this point a weaker 
hierarchy between $\langle E_{x} \rangle$ and $\langle E_{\bar e}
\rangle$ is considered, being about 10\% for point 3 and about 20\%
for point 1. The main difference between these points, is the fact
that in the case i0, the hierarchy can no longer be recognized.  
As mentioned above, the establishment of the hierarchy dependents 
on whether a value of $\langle E_{\bar e} \rangle$ equal to 
$\langle E_{x} \rangle$ is included in the scan over parameters. 
Had we enlarged the SN parameter space in Eq.~(\ref{snparamarea}) 
to include $\langle E_{\bar e} \rangle=18$ MeV we would find a 
very small allowed area for the WaterC detector for the normal 
hierarchy around the set of values: 
$\langle E_{\bar e} \rangle=18$ MeV, 
$\langle E_{e} \rangle \simeq 10$ MeV, 
$\langle E_{x} \rangle \simeq 18$ MeV, $\xi \simeq 0.8$,
$\sin^2\theta_{13} \simeq 10^{-4}$ and $E_b/D^2 \simeq 3.1 \times
10^{51}$ ergs/kpc$^2$.  It is not difficult to realize that this
set of parameters produces total number of events in each 
channel only slightly different from the i0 case. Improving 
the energy resolution of the detector does not help much. 
For point 1 the normal hierarchy 
is excluded at 95\% CL and for point 3 by less than 1$\sigma$ for 
the WaterC detector. Nevertheless, as the allowed region for the 
normal hierarchy in this case is very small and occurring for the 
$\bar \nu_e$ and $\nu_x$ fluxes having equal temperatures, 
one would naturally have a strong hint that the true hierarchy 
is indeed the inverted hierarchy. 
An analog discussion could be performed for the case n0. However, as 
this would require all three average energies to be almost equal, 
we will refrain from this discussion, as it is physically very 
improbable.

Next, we would like to discuss how the detection of Earth matter
effects and shock wave effects can help to pin down the 
hierarchy in the i0 scenario when allowing the $\bar \nu_e$ 
and $\nu_x$ fluxes to have identical temperatures.  
Let us shortly review the facts about Earth matter effects for 
SN neutrinos~\cite{Lunardini:2001pb,Dighe:2003vm}.
For neutrinos that traverse the Earth mantle (and core) a modulation
with known frequencies of the neutrino energy spectra 
may occur, depending on the hierarchy and the value of $\theta_{13}$. 
If the hierarchy is normal (inverted), the Earth matter effect 
for neutrinos (anti-neutrinos) depend on the value of $P_H$. 
In both cases, the strength of the 
modulation will be proportional to the difference in the original 
$\nu_x$ and $\nu_e$ ($\nu_{\bar e}$) fluxes and the value of $P_H$. 
Therefore, in the case of normal hierarchy and large values of 
$\theta_{13}$ ($P_H \simeq 0)$) there should be a modulation of 
the anti-neutrino energy-spectra {\it unless} 
$\langle E_{\bar e} \rangle= \langle E_{x} \rangle$ and no 
modulation of the neutrino spectra.
Similarly, in the case i0 there should be a modulation of the neutrino
energy-spectra {\it unless} $\langle E_{e} \rangle= \langle E_{x}
\rangle$ and there should be no modulation of the anti-neutrino
spectra.  This can clearly help to distinguish the hierarchy in case
i0 for point 1, as we do have a hierarchy between the neutrino and 
$\nu_x$ average energies. With detectors shielded and unshielded by 
Earth, telling us that there are Earth matter effects in the 
neutrino channel and not in the anti-neutrino channel, the case 
i0 is clearly established.
The expectation for point 1, is a maximum difference of the 
neutrino flux arriving at Earth and the flux after traversing 
part of the Earth matter of about 20\%, occurring for a neutrino 
energy of roughly 60 MeV. Obviously for point 3 it will be 
much more difficult to observe and determine the Earth matter 
effects, as the weak hierarchy between $\langle E_{\bar e} \rangle$ 
and $\langle E_{x} \rangle$ will make the overall strength 
of the Earth matter effects smaller. Furthermore, shock-wave 
effects~\cite{shock} may also help to identify the 
true hierarchy. These effects can be seen in the adiabatic region 
of the high (H) resonance, {\it i.e.} for large values of 
$\theta_{13}$. In the case of inverted hierarchy a dip in the 
value of $\langle E_{\bar e} \rangle$ as a function of time is 
expected, whereas for normal hierarchy the dip is expected for 
$\langle E_{e} \rangle$. Therefore, an observation of shock 
wave effects in the anti-neutrino channel will point toward 
the inverted hierarchy. Once again the 
amplitude of this effects decrease as the $\nu_e$/$\bar \nu_e$ 
and $\nu_x$ temperatures becomes closer, making it difficult to 
pin down the hierarchy for point 1.  
In conclusion, the hierarchy for case i0 is likely to be  
established with the detection of the SN neutrinos at Earth 
alone. The confidence level with which this can be done increase 
as the hierarchy between the $\bar \nu_e$ and $\nu_x$ temperatures 
increase. With the complementary information on Earth matter effects 
and shock wave effects it is very likely that the hierarchy can be 
undoubtedly established for point 1.

Summarizing, the values of some or even all of the SN parameters as 
well as the unknown neutrino ones, might be determined 
simultaneously, in most cases. The values of $\xi$, 
$\langle E_{\bar e} \rangle$ and $\langle E_{e} \rangle$ are 
however difficult to determine with a high precision. 
Overall, there is not much difference in the performance of 
all three detectors. The WaterC 
can access most SN parameters with a higher accuracy, but this is 
basically due to the larger mass and hence statistics. The LENA-type 
detector is performing slightly better than LAr (except for the 
determination of $\langle E_e \rangle$) for the cases a1 and i0. 
For the n0 case they are almost equally good, with LAr determining 
$\langle E_e \rangle, \langle E_x \rangle$ better and 
LENA doing a better job in determining $\xi$. Therefore, 
generally the two detectors have similar performances, although 
the mass of the LAr detector is twice as big as that of LENA. 
This difference can be understood as the LAr detector has a 
rather weak sensitivity to the pure anti-electron neutrino flux. 
The LENA-type detector, on the other hand, has a good distribution 
between sensitivity to pure $\nu_e$, pure $\bar \nu_e$ and the 
total neutrino flux. 

\section{Conclusions}
\label{sec:conclusions}

We have studied the prospects for extracting, simultaneously SN
parameters and neutrino oscillation parameters from the measurements
of neutrinos from the cooling phase of a galactic supernova in three
different detectors: a megaton-scale water Cherenkov, a 100 kton
liquid Argon and a 50 kton scintillation detector.

In our analysis we have varied a total of seven parameters, 
five SN parameters: the average energies $\langle E_{e}\rangle$, 
$\langle E_{\bar e}\rangle$ and $\langle E_{x}\rangle$, 
the ratio of the luminosities $\xi$ and the overall normalization 
of the fluxes $E_b/D^2$; two neutrino oscillation parameters: the 
angle $\theta_{13}$ and the neutrino mass hierarchy.
Since we considered perfect detectors, with 100\% efficiencies, our 
analysis must be viewed as an estimation of the maximal performance of 
each experimental setup.
We do not include Earth matter effects or shock-wave effects 
in our calculations but briefly discuss their possible implications.

Our main results are summarized in Figs. \ref{fig:hki1}-\ref{fig:hkn0},
\ref{fig:lai1}-\ref{fig:lan0} and \ref{fig:sci1}-\ref{fig:scn0}, for 
the WaterC, LAr and Scintillator detectors, respectively.
We have found that SN parameters, as well as the unknown neutrino ones,
can be determined simultaneously, in most cases. 
Comparing the three detectors, there is not much difference in 
their overall performance.  However, the WaterC detector can access 
most SN parameters with higher accuracy (a few \% in some cases), 
but this is basically due to its larger mass and statistics.  

All of the studied detectors have the possibility to determine the 
neutrino mass hierarchy if 
$\sin^2(\theta_{13}) \stackrel{>}{\sim} 10^{-4}$ and 
the hierarchy of the average energies is stronger than about 20\%. 
They can at the same time determine some of the SN parameters 
quite well and put strong bounds on the value of $\theta_{13}$. 
The average energy of the $\nu_\mu$ and $\nu_\tau$ species can be 
determined with an accuracy better than 5\% in most of the 
parameter space suggested by SN simulations. 

Although the detection of several separable channels measuring
different combinations of CC and NC processes is crucial for the
determination of $\theta_{13}$ and the hierarchy, there are cases
where some SN parameters can be determined rather well 
even when only the main CC detection channel is available.


   
\section*{ \label{APP:nonopt} APPENDIX A: Allowed regions for 
detectors at non-optimal performances}

In this appendix we show contours for a few cases, where we 
assume that some channels are either non-separable from 
other channels or not available at all.

\begin{figure}[p]
\centering
\includegraphics[width=17cm,height=19cm]{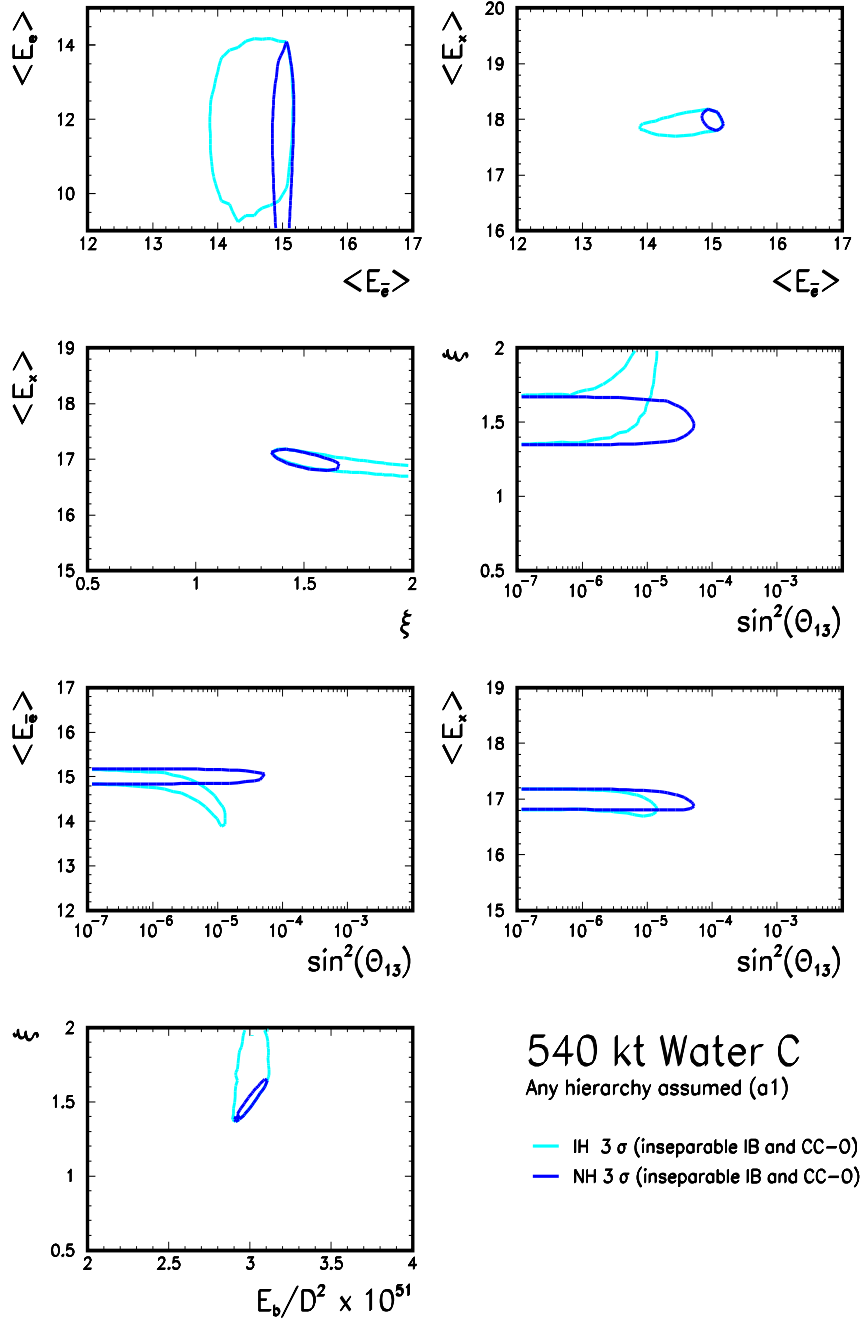}
\caption{Sensitivity of a 540 kton WaterC detector assuming true 
SN parameters as in point 1 of Table \ref{tab:points}, 
for any hierarchy and $\sin^2(\theta_{13})=10^{-6}$ ($P_H \simeq 1$).
We show  3$\sigma$ CL contours (2 dof) using all 4 detection 
processes (IB + CC-O + ELAS+ NC-O), but assuming that the IB and 
CC-O cannot be distinguished. The contours for the normal 
hierarchy (NH) is marked by the dark (red) and for the inverted 
hierarchy (IH) is marked by the light (cyan) solid line.}
\label{fig:hki1-c2}
\end{figure}
\begin{figure}[p]
\centering
\includegraphics[width=17cm,height=19cm]{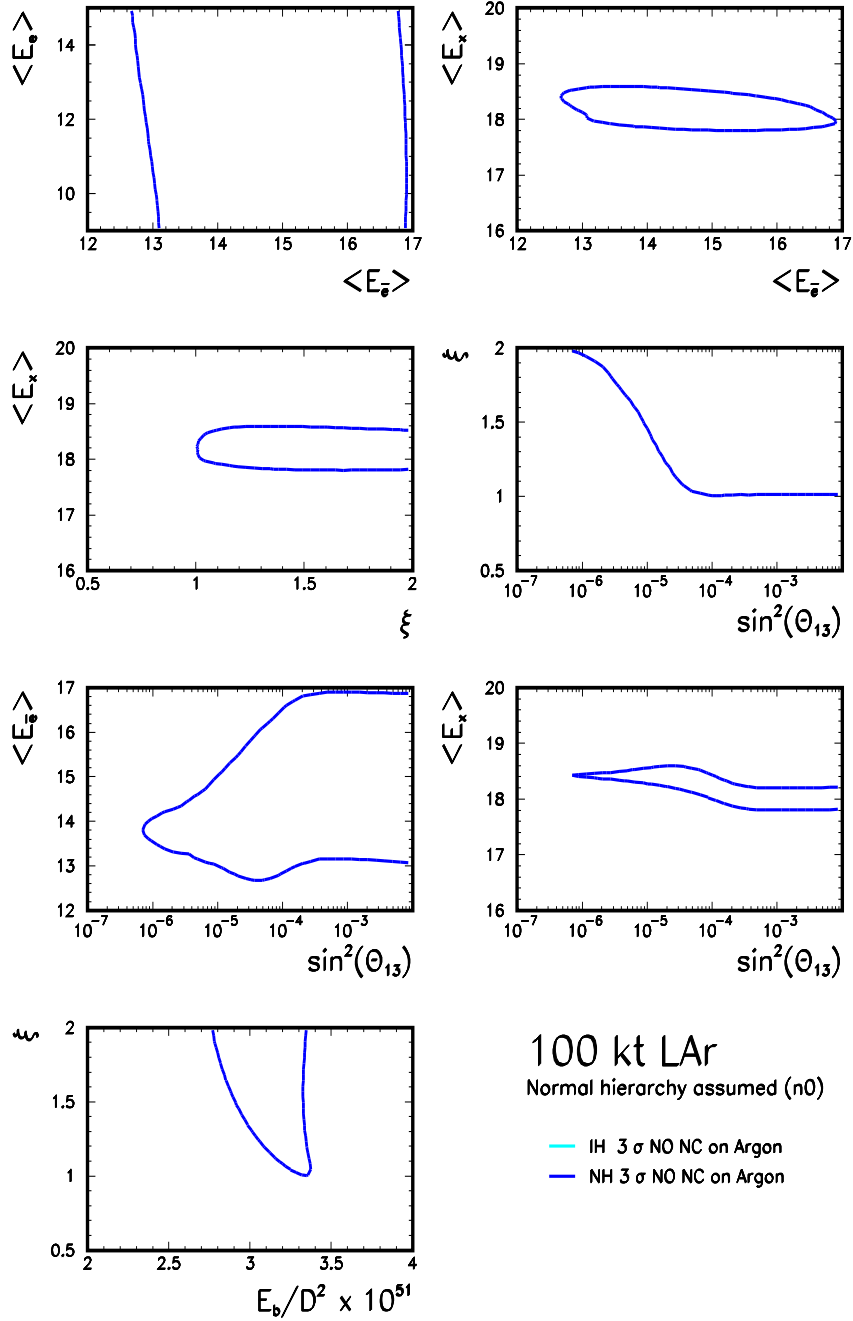}
\caption{Sensitivity of a 100 kton LAr detector assuming true 
SN parameters as in point 1 of Table \ref{tab:points}, for 
normal hierarchy and $\sin^2(\theta_{13})=10^{-3}$ ($P_H \simeq 0$).
We show  3$\sigma$ CL contours (2 dof) using 3 channels  
($\nu_e$CC+ $\bar\nu_e$CC + ELAS), assuming that the NC on 
Argon is not available.  
The contours for normal hierarchy is marked by the dark (red) 
and for inverted hierarchy is marked by the light (cyan) solid line.
The inverted hierarchy is ruled out by more than 5$\sigma$. }
\label{fig:lan0-c2}
\end{figure}
\begin{figure}[p]
\centering
\includegraphics[width=17cm,height=19cm]{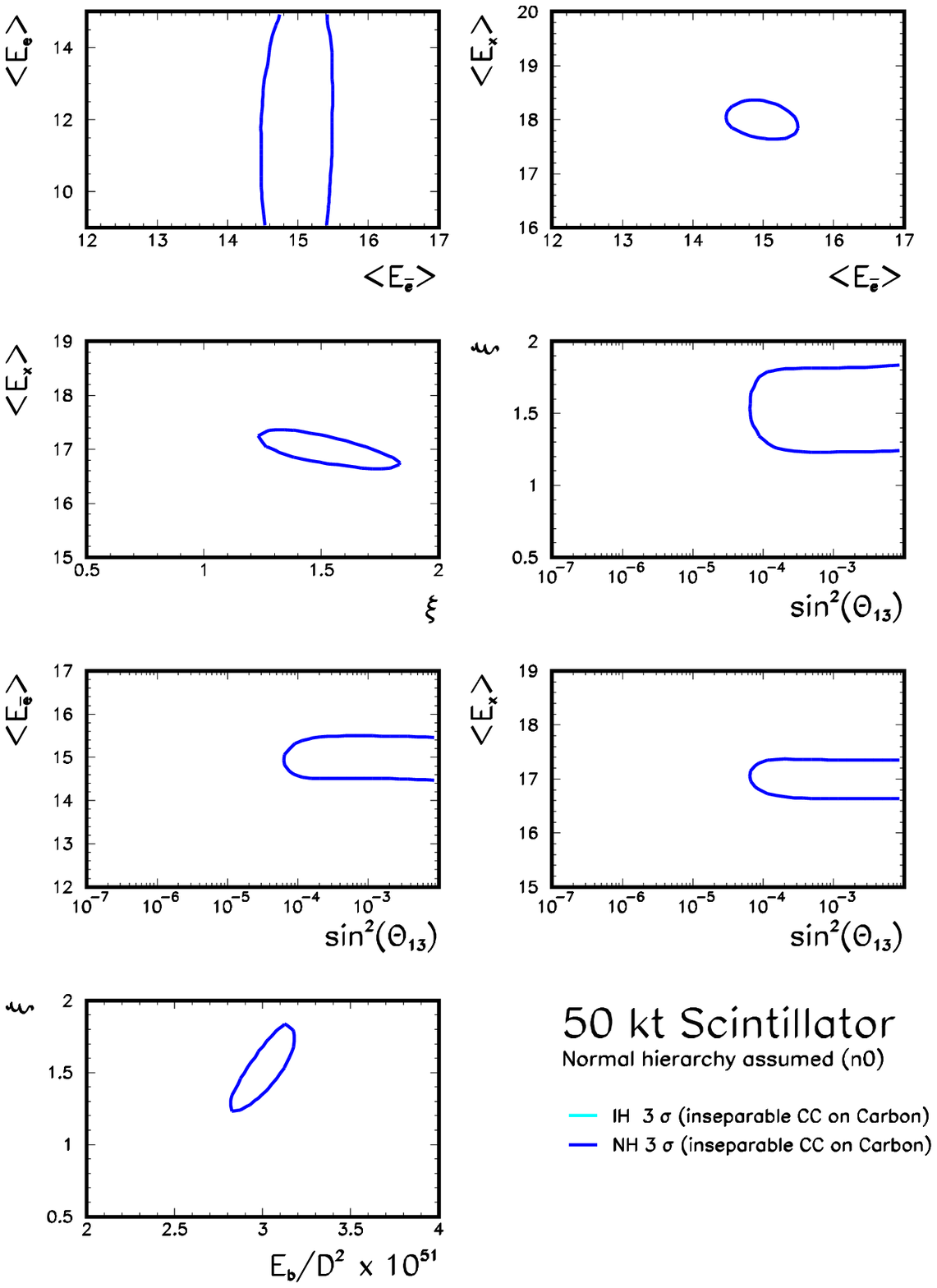}
\caption{Sensitivity of a 50 kton Scintillation detector assuming 
true SN parameters as in point 1 of Table \ref{tab:points}, for 
normal hierarchy and  $\sin^2(\theta_{13})= 10^{-3}$  ($P_H \simeq 0$).
We show  3$\sigma$ CL contours (2 dof) using all 6 detection
processes (IB + $\nu_e$CC+ $\bar\nu_e$CC + $\nu-p$ + NC + ELAS) but 
assuming that the  $\nu_e$CC and $\bar\nu_e$CC cannot be
distinguished. The contours for normal hierarchy 
is marked by the dark (red) and for inverted hierarchy is marked 
by the light (cyan) solid line. 
The inverted hierarchy is ruled out by more than 5$\sigma$. 
}  
\label{fig:scn0-c2}
\end{figure}

We investigate the case of a WaterC detector where 
the inverse beta decay channel cannot be separated from the 
CC on Oxygen. We have chosen this scenario, as indeed this 
separation might be difficult if gadolinium will not be 
added (in which case the neutron is not detectable), 
since both detection processes are almost isotropic. 
We show the a1 case in Fig.\ref{fig:hki1-c2}. 
Comparing to the case of all channels being separable in 
Fig.\ref{fig:hki1} (solid lines) we see that the biggest difference is the 
determination of $\sin^2\theta_{13}$ for the normal hierarchy 
allowed region. The upper bound on $\sin^2\theta_{13}$ increases 
from $2 \times 10^{-5}$ to $5 \times 10^{-5}$. 
Besides this the other parameters have very
similar restrictions. Moreover, the contours for inverted hierarchy 
are almost identical to the ones in Fig.\ref{fig:hki1} (solid lines). 
This is expected as for the inverted hierarchy the inverse beta decay 
channel dominates and therefore whether or not it can be separated from 
the CC-O does not have a large effect. This is also the reason 
that the scenario i0 does not change much if the IB and CC-O cannot 
be separated. The only visible difference is a slightly worse
determination of $\langle E_e \rangle$, but this variable can not 
be determined very well in either scenarios. For the scenario n0, the 
lower bound on $\sin^2\theta_{13}$ jumps from $1 \times 10^{-4}$ to 
$6 \times 10^{-5}$. 
Overall, the separation of the IB and CC-O channel have little impact 
on the detector performance, besides a less restrictive bound on the 
CHOOZ angle.

Next we study the LAr detector in the case that the neutral current 
on Argon is not available. This NC channel is not accompanied by the 
detection of the positron or electron and might therefore be difficult 
to trigger, especially if the detector is not well shielded. 
We show the case where the input scenario is n0 in Fig.\ref{fig:lan0-c2}. 
Comparing Fig.\ref{fig:lan0-c2} to Fig.\ref{fig:lan0} (solid lines) 
we see that again the biggest impact is a worsening of the 
determination of $\sin^2\theta_{13}$. In this case the impact is 
severe, lowering the upper bound from $10^{-4}$ to $7 \times
10^{-7}$. Other parameters are only mildly affected. 
For the case i0, the normal hierarchy cannot be ruled out with 
the 3 channel scenario, as a small 'fake' region around 
$\langle E_{\bar e} \rangle = 17$ MeV is allowed for small values 
of $\sin^2\theta_{13}$. 
This also leaves $\sin^2\theta_{13}$ unrestricted for the inverted
hierarchy, unless one constrains $\langle E_{\bar e} \rangle$ to 
be less than about 16.5 MeV, in which case the restriction is 
roughly as the case with 4 channels. 
For the input scenario a1, again the biggest impact is a drop in 
the upper bound on  $\sin^2\theta_{13}$ from $2 \times 10^{-5}$ 
to $6 \times 10^{-5}$. 
We conclude that the NC channel on Argon 
is an important factor for determining the neutrino parameters:
the hierarchy and in particular the value of the CHOOZ angle. 

Finally we investigate the Scintillator detector assuming
that the neutrino and anti-neutrino CC reactions on Carbon cannot 
be separated. 
We show the case of the input scenario n0 
in Fig.\ref{fig:scn0-c2}. 
Comparing Fig.\ref{fig:scn0-c2} to Fig.\ref{fig:scn0} (solid lines) 
we see that there is hardly any visible difference. This is also 
valid for the two other cases (a1 and i0). The reason for this is 
that the dominant inverse beta decay channel is still a separate 
channel and thus gives good restrictions on the $\bar \nu_e$ flux, 
which in turn makes a de facto separation of the $\nu_e$-CC and 
$\bar \nu_e$-CC channels when doing the fitting.  
We conclude that the separation of the CC channels on Carbon 
only have a minor impact on the performance of the Scintillator
detector.

\begin{acknowledgments}
We thank Funda\c{c}\~ao de Amparo \`a Pesquisa do Estado de S\~ao Paulo
(FAPESP) and Conselho Nacional  de Ci\^encia e Tecnologia (CNPq) for 
financial support.
\end{acknowledgments}


\end{document}